%% file: ms.tex
\RequirePackage{amsmath} % load amsmath first to avoid \vec redefinition warning
\documentclass[twocolumn]{svjour3}
\pdfoutput=1

\usepackage{graphicx}
\usepackage[T1]{fontenc}
\usepackage[utf8]{inputenc}

\usepackage{newtxtext,newtxmath}
\usepackage{multirow}
\usepackage{subcaption}
\usepackage{enumitem}
\usepackage{balance}
\usepackage{url}

\usepackage{hyphenat}
\usepackage{pdfcomment}
\usepackage{booktabs}
\usepackage{verbatim}
\usepackage{tikz}
\usetikzlibrary{arrows}
\usepackage{pdfpages}
\usepackage{color}
\usepackage{times}
\usepackage[nocompress]{cite}
\let\subparagraph\section
\usepackage[compact]{titlesec}
\usepackage{hyperref}

\smartqed% flush right qed marks, e.g. at end of proof
\PassOptionsToPackage{unicode}{hyperref}
\PassOptionsToPackage{naturalnames}{hyperref}

\newif\iftablesqueeze{}
\tablesqueezetrue{}

\renewcommand{\theenumi}{\roman{enumi}}

\newcommand{\customheader}[1]{\vspace{0.2cm}\noindent\emph{#1}}

\newenvironment{squishedenumerate}
{
  \begin{enumerate}[label=(\roman*),topsep=0.2em,leftmargin=2em]
}
{
  \end{enumerate}
}

\newenvironment{squishedlist}
{
  \begin{list}{$\bullet$}
  {
    \setlength{\itemsep}{0pt}
    \setlength{\parsep}{2pt}
    \setlength{\topsep}{0.3em}
    \setlength{\partopsep}{0pt}
    \setlength{\leftmargin}{1.5em}
    \setlength{\labelwidth}{1em}
    \setlength{\labelsep}{0.5em}
  }
}
{
  \end{list}
}

\newenvironment{squishedtable}
{
  \begin{table}[t]
}
{
  \end{table}
}

\newenvironment{squishedtablestar}
{
  \begin{table*}[t]
}
{
  \end{table*}
}

\journalname{}

\begin{document}

\title{The Ubiquity of Large Graphs and Surprising Challenges of Graph Processing: Extended Survey}

\author{Siddhartha Sahu \and Amine Mhedhbi \and
    Semih Salihoglu \and Jimmy Lin \and M. Tamer \"Ozsu
}

\institute{Siddhartha Sahu, Amine Mhedhbi, Semih Salihoglu, Jimmy Lin, and M. Tamer \"Ozsu
    \at{}
    University of Waterloo \\
    \email{\{s3sahu, amine.mhedhbi, semih.salihoglu, jimmylin, tamer.ozsu\}@uwaterloo.ca}
}

\date{}

\maketitle

\input{00abstract}

\input{01introduction}

\input{02methodology}
\input{03survey}

\input{04applicationswhitepapers}

\input{05applicationsinterviews}

\input{06endtext}

\bibliographystyle{spmpsci}
\bibliography{references}

\input{07appendix}

\end{document}

%% file: 00abstract.tex
\begin{abstract}

    Graph processing is becoming increasingly prevalent across many application
    domains. In spite of this prevalence, there is little research about how
    graphs are actually used in practice. We performed an extensive study that
    consisted of an online survey of 89 users, a review of the mailing lists, source
    repositories, and whitepapers of a large suite of graph software products, 
    and in-person interviews with 6 users and 2 developers of these products.
    Our online survey aimed at understanding: (i) the types of graphs users have; 
    (ii) the graph computations users run; (iii) the types of graph software users use; and
    (iv) the major challenges users face when processing their graphs. We
    describe the participants' responses to our questions highlighting common
    patterns and challenges. Based on our interviews and survey of the rest of our sources,
    we were able to answer some new questions that were raised by participants'
    responses to our online survey and understand the specific applications 
    that use graph data and software.
    Our study revealed surprising facts about graph processing 
    in practice. In
    particular, real-world graphs represent a very diverse range of entities and
    are often very large, scalability and visualization are undeniably the
    most pressing challenges faced by participants, and data integration, recommendations, and fraud detection are very popular applications supported by existing graph software.
    We hope these findings can guide future research.
\end{abstract}

%% file: 01introduction.tex
\section{Introduction}

\noindent Graph data representing connected entities and their relationships
appear in many application domains, most naturally in social networks, the web,
the semantic web, road maps, communication networks, biology, and finance, just
to name a few examples. There has been a noticeable increase in the prevalence
of work on graph processing both in research and in practice, evidenced by the
surge in the number of different commercial and research software for managing
and processing graphs. Examples include graph database
systems~\cite{neo4j,orientdb,arrangodb,dgraph,janusgraph,caley,sparksee}, RDF
engines~\cite{jena,virtuoso}, linear algebra
software~\cite{blas,matlab}, visualization
software~\cite{cytoscape,elasticsearch-graph}, query
languages~\cite{gremlin,opencypher,pgql}, and distributed graph processing
systems~\cite{giraph,graphx,flink}. In the academic literature, a large number
of publications that study numerous topics related to graph processing regularly
appear across a wide spectrum of research venues.

Despite their prevalence, there is little research on how graph data is actually
used in practice and the major challenges facing users of graph data, both in
industry and research. In April 2017, we conducted an online survey across 89
users of 22 different software products, with the goal of answering 4 high-level
questions:

\begin{squishedenumerate}
    \item What types of graph data do users have?
    \item What computations do users run on their graphs?
    \item Which software do users use to perform their computations?
    \item What are the major challenges users face when processing their graph
    data?
\end{squishedenumerate}

\noindent Our major findings are as follows:

\begin{squishedlist}

    \item \emph{Variety:} Graphs in practice represent a very wide variety of
    entities, many of which are not naturally thought of as vertices and edges.
    Most surprisingly, traditional enterprise data comprised of products,
    orders, and transactions, which are typically seen as the perfect fit for
    relational systems, appear to be a very common form of data represented in
    participants' graphs.

    \item \emph{Ubiquity of Very Large Graphs:} Many graphs in practice are very
    large, often containing over a billion edges. These large graphs represent a
    very wide range of entities and belong to organizations at all scales from
    very small enterprises to very large ones. This refutes the sometimes heard
    assumption that large graphs are a problem for only a few large
    organizations such as Google, Facebook, and Twitter.

    \item \emph{Challenge of Scalability:} Scalability is unequivocally the most
    pressing challenge faced by participants. The ability to process very large
    graphs efficiently seems to be the biggest limitation of existing software.

    \item \emph{Visualization:} Visualization is a very popular and central task
    in participants' graph processing pipelines. After scalability, participants
    indicated visualization as their second most pressing challenge, tied with
    challenges in graph query languages.

    \item \emph{Prevalence of RDBMSes:} Relational databases still play an
    important role in managing and processing graphs.

\end{squishedlist}

\noindent Our survey also highlights other interesting facts, such as the
prevalence of machine learning on graph data, e.g., for clustering vertices,
predicting links, and finding influential vertices.

We further reviewed user feedback in the mailing lists, bug reports, and feature
requests in the source code repositories of 22 software products between January
and September of 2017 with two goals: (i)~to answer several new questions that
the participants' responses raised; and (ii)~to identify more specific
challenges in different classes of graph technologies than the ones we could
identify in participants' responses. For some of the questions in our online
survey, we also compared the graph data, computations, and software used by the
participants with those studied in academic publications. For this, we reviewed
252 papers from 3 different year's proceedings of 7 conferences across different academic venues.

Different database technologies and research topics are often motivated with a small set of common applications, informally referred to as ``killer'' applications of the technology. For example, object-oriented database systems are associated with computer-aided design and manufacturing, and XML is associated with the web. An often-asked question in the context of graphs is: What is the killer application of graph software products? The wide variety of graphs and industry fields mentioned by our online survey participants hinted that we cannot pinpoint a small set of such applications.
To better understand the applications supported by graphs, we reviewed the whitepapers posted on the websites of 8 graph software products. We also interviewed 6 users and 2 developers of graph processing systems. Our reviews and interviews corroborated our findings that graphs have a very wide range of applications but also highlighted several common applications, primarily in data integration, recommendations, and fraud detection, as well as several new applications we had not identified in our online survey. Our interviews also give more details than our online survey about the actual graphs used by enterprises and how they are used in applications.

In addition to discussing the insights we gained through our study, we discuss
several directions about the future of graph processing. We hope our study can
inform research about real use cases and important problems in graph processing.

%% file: 02methodology.tex
\section{Methodology of Online Survey, Mailing Lists, Source Repositories, and Academic Publications}\label{sec:methodology}
In this section, we first describe the format of our survey and then
how we recruited the participants. Next we describe the demographic information of the participants, including the organizations
they come from and their roles in their organizations. Then we
describe our methodology of reviewing academic publications. Then we 
describe our methodology for reviewing the
user feedback in the mailing lists, bug reports, and feature requests
in the source code repositories of the software products. We end this section with a discussion of our methodology, which we believe other
       researchers can easily reproduce to study the uses of other technology, and some lessons we learned from our experience of performing a user study. We review our methodology of reviewing whitepapers and our interviews in Sections~\ref{subsec:whitepaper-methodology} and~\ref{subsec:interview-methodology}, respectively.

\subsection{Online Survey Format and Participants}%
\label{subsec:online-survey}
\subsubsection{Format}
\noindent The survey was in the format of an online form. All of the questions
were optional and participants could skip any number of questions. There were 2
types of questions:

\begin{squishedenumerate}

    \item \textit{Multiple Choice}: There were 3 types of multiple choice
    questions: (a)~yes-no questions; (b)~questions that allowed only a single
    choice as a response; and (c)~questions that allowed multiple choices as a
    response. The participants could use an \emph{Other} option when their
    answers required further explanation or did not match any of the provided
    choices. We randomized the order of choices in questions about the
    computations participants run and the challenges they face.

    \item \textit{Short Answer}: For these questions, the participants entered
    their responses in a text box.

\end{squishedenumerate}

\noindent There were 34 questions grouped into six categories: (i)~demographic
questions; (ii)~graph datasets; (iii)~graph and machine learning computations;
(iv)~graph software; (v) major challenges; and (vi)~workload breakdown.

\subsubsection{Participant Recruitment}\label{sec:participant-recruitment}
\noindent We prepared a list of 22 popular software products for processing
graphs (see Table~\ref{table:graph-software}) that had public user mailing lists
covering 6 types of technologies: graph database systems, RDF engines,
distributed graph processing systems (DGPSes), graph libraries to run and
compose graph algorithms, visualization software, and graph query
languages.\footnote{The linear algebra software we considered, e.g.,
    BLAS~\cite{blas} and MATLAB~\cite{matlab}, either did not have a public
    mailing list or their lists were inactive.} Our goal was to be as
comprehensive as possible in recruiting participants from the users of different
graph technologies. However, we acknowledge that this list is incomplete and
does not cover all of the graph software used in practice.

\begin{squishedtable}
    \centering \setlength{\tabcolsep}{0.4em}
    \caption{Software products used for recruiting participants and the count of
        active users in their mailing list in Feb-Apr 2017. Last column is the 
        total count of each type of software.}%
    \label{table:graph-software}
    \iftablesqueeze\vspace{-0.7em}\fi
    \begin{tabular}{p{2.55cm}lrr}
      \toprule
      \textbf{Technology}                                                  & \textbf{Software}                                    & \multicolumn{2}{l}{\textbf{\# Users}} \\ \midrule
      \multirow{7}{*}{\parbox{1.9cm}{Graph Database System}}               & ArangoDB\cite{arrangodb}                             & 40 & \multirow{7}{*}{238}             \\
                                                                           & Caley\cite{caley}                                    & 14 &                                  \\
                                                                           & DGraph\cite{dgraph}                                  & 33 &                                  \\
                                                                           & JanusGraph\cite{janusgraph}                          & 32 &                                  \\
                                                                           & Neo4j\cite{neo4j}                                    & 69 &                                  \\
                                                                           & OrientDB\cite{orientdb}                              & 45 &                                  \\
                                                                           & Sparksee\cite{sparksee}                              & 5  &                                  \\ \midrule
      \multirow{2}{*}{RDF Engine}                                          & Apache Jena\cite{jena}                               & 87 & \multirow{2}{*}{110}             \\
                                                                           & Virtuoso\cite{virtuoso}                              & 23 &                                  \\ \midrule
      \multirow{3}{*}{\parbox{2.2cm}{Distributed Graph Processing Engine}} & Apache Flink (Gelly)\cite{flink}                     & 24 & \multirow{3}{*}{39}              \\
                                                                           & Apache Giraph\cite{giraph}                           & 8  &                                  \\
                                                                           & Apache Spark (GraphX)\cite{graphx}                   & 7  &                                  \\ \midrule
      Query Language                                                       & Gremlin\cite{gremlin}                                & 82 & 82                               \\ \midrule
      \multirow{6}{*}{Graph Library}                                       & Graph for Scala\cite{graph-for-scala}                & 4  & \multirow{6}{*}{97}              \\
                                                                           & GraphStream\cite{graphstream}                        & 8  &                                  \\
                                                                           & Graphtool\cite{graphtool}                            & 28 &                                  \\
                                                                           & NetworKit\cite{networkit}                            & 10 &                                  \\
                                                                           & NetworkX\cite{networkx}                              & 27 &                                  \\
                                                                           & SNAP\cite{snap}                                      & 20 &                                  \\ \midrule
      \multirow{2}{*}{Graph Visualization}                                 & Cytoscape\cite{cytoscape}                            & 93 & \multirow{2}{*}{116}             \\
                                                                           & Elasticsearch X-Pack Graph\cite{elasticsearch-graph} & 23 &                                  \\ \midrule
      \parbox{2.8cm}{Graph Representation}                                 & Conceptual Graphs\cite{conceptualgraphs}             & 6  & 6                                \\ \bottomrule
    \end{tabular}
\end{squishedtable}

We conducted the survey in April 2017, and used 4 methods to recruit
participants from the users of these 22 software products:
\begin{squishedlist}

    \item \textit{Mailing Lists}: We posted the survey to the user mailing lists
    of the software in our list.

    \item \textit{Private Emails}: Five mailing lists:\ (i)~Neo4j;
    (ii)~OrientDB;\@ (iii)~ArangoDB;\@ (iv)~JanusGraph; and (v)~NetworkX,
    allowed us to send private emails to the users. We sent private emails to
    171 users who were active on these mailing lists between February and April
    of 2017.

    \item \textit{Slack Channels}: Two of the software products on our list,
    Neo4j and Cayley, had Slack channels for their users. We posted the survey
    to these channels.

    \item \textit{Twitter}: A week after posting our survey to the mailing lists
    and Slack channels and sending private emails, we posted a tweet with a link
    to our survey to 7 of the 22 software products that had an official Twitter
    account. Only Neo4j retweeted our tweet.

\end{squishedlist}

\noindent Participants that we recruited through different methods shared the
same online link and we could not tell the number of participants recruited from
each method. In particular, we suspect that there were more users from graph
database systems mainly because their mailing lists contained more active users,
as can be seen in Table~\ref{table:graph-software}. Moreover, 4 of the 5 mailing
lists that allowed us to send private emails and the Slack and Twitter channels
belonged to graph database systems. We note that after posting the survey on
Twitter, we received 12 responses.

In the end, there were 89 participants. Below, we give an overview of the
organizations these participants work in and the role of the participants in
their organizations.

\vspace{0.5em} \noindent \textbf{Field of Organizations}: We asked the
participants which field they work in. Participants could select multiple
options. Table~\ref{table:field-of-work} shows the 12 choices and participants'
responses. In the table, ``R'' and ``P'' indicate researchers and practitioners
(defined momentarily), respectively. In addition to the given choices, using the
\emph{Other} option, participants indicated 5 other fields: education, energy
market, games and entertainment, investigations and audits, and grassland
management. In total, participants indicated 17 different fields, demonstrating
that graphs are being used in a wide variety of fields. Throughout the survey,
we group the participants into 2 categories:

\begin{squishedtable}
    \centering
    \caption{The participants' fields of work.}%
    \label{table:field-of-work}
    \iftablesqueeze\vspace{-0.7em}\fi
    \begin{tabular}{p{3.7cm}rrr}
      \toprule
      \multicolumn{1}{l}{\textbf{Field}} & \textbf{Total} & \textbf{R} & \textbf{P} \\ \midrule
      Information \& Technology          & 48             & 12         & 36         \\
      Research in Academia               & 31             & 31         & 0          \\
      Finance                            & 12             & 2          & 10         \\
      Research in Industry Lab           & 11             & 11         & 0          \\
      Government                         & 7              & 3          & 4          \\
      Healthcare                         & 5              & 3          & 2          \\
      Defense \& Space                   & 4              & 3          & 1          \\
      Pharmaceutical                     & 3              & 0          & 3          \\
      Retail \& E-Commerce               & 3              & 0          & 3          \\
      Transportation                     & 2              & 0          & 2          \\
      Telecommunications                 & 1              & 1          & 0          \\
      Insurance                          & 0              & 0          & 0          \\
      Other                              & 5              & 2          & 3          \\
      \bottomrule
    \end{tabular}
\end{squishedtable}

\begin{squishedlist}

    \item \emph{Researchers} are the 36 participants who indicated at least one
    of their fields as research in academia or research in an industry lab. Some
    of these participants further selected other choices as their fields, the
    most popular of which were information and technology, government, defense
    and space, and health care.

    \item \emph{Practitioners} are the remaining 53 participants who did not
    select research in academia or an industry lab. The top two fields of
    practitioners were information and technology and finance, indicated by 36
    and 10 people, respectively.

\end{squishedlist}

\noindent In the remainder of this paper, we will explicitly indicate when the
responses of the researchers and practitioners to our survey questions differ
significantly. In the absence of an explicit comparison, readers can assume that
both groups' responses were similar.

\vspace{0.5em} \noindent \textbf{Size of Organizations}:
Table~\ref{table:size-of-organization} shows the sizes of the organizations that
the participants work in, which ranged from very small organizations with less
than 10 employees to very large ones with more than 10,000 employees.

\begin{squishedtable}
    \centering
    \caption{Size of the participants' organizations.}%
    \label{table:size-of-organization}
    \iftablesqueeze\vspace{-0.7em}\fi
    \begin{tabular}{p{3.7cm}rrr}
      \toprule
      \multicolumn{1}{l}{\textbf{Size}} & \textbf{Total} & \textbf{R} & \textbf{P} \\ \midrule
      1$-$10                            & 27             & 17         & 10         \\
      10$-$100                          & 23             & 6          & 17         \\
      100$-$1000                        & 14             & 4          & 10         \\
      1000$-$10000                      & 6              & 4          & 2          \\
      \textgreater{} 10000              & 15             & 4          & 11         \\
      \bottomrule
    \end{tabular}
\end{squishedtable}

\vspace{0.5em} \noindent \textbf{Role at Work}: We asked the participants their
roles in their organizations and gave them the following 4 choices:
(i)~researcher; (ii)~engineer; (iii)~manager; and (iv)~data analyst.
Participants could select multiple options. The top 4 roles were engineers,
selected by 54, researchers, selected by 48, data analysts, selected by 18, and
managers, selected by 16. The other roles participants indicated were architect,
devops, and student.

\subsection{Review of Academic Publications}%
\label{sec:academic-publications-review}

\noindent In order to compare the graph data, computations, and software
academics work on with those that our participants indicated, we surveyed papers
in the proceedings of 3~different years of the 7~academic conferences shown in
Table~\ref{table:conferences}.\footnote{For each conference, we
        initially surveyed one year selected between 2014 to 2016 and later
        extended the survey to include the years 2017 and 2018. Note that OSDI
        and SOSP are held in alternating years.} Our goal in choosing these
conferences was to select a variety of venues where
papers on graph processing are published. Specifically, our list consists of venues
in databases, data mining, machine learning, operating systems, high performance
computing, and cloud computing. For each paper in these proceedings, we first
selected the ones that directly studied a graph computation or were developing
graph processing software. We omitted papers that were not primarily focused on
graph processing, even if they used a graph algorithm as a subroutine to solve a
problem. For example, we omitted a paper studying a string matching algorithm
that uses a graph algorithm as a subroutine. In the end, we selected 252 papers.

\begin{squishedtable}
    \centering
    \caption{Academic conferences and surveyed years.}%
    \label{table:conferences}
    \iftablesqueeze\vspace{-0.7em}\fi
    \begin{tabular}{ll}
      \toprule
      \textbf{Conference} & \textbf{Years reviewed}                                          \\ \midrule
      VLDB                & 2014~\cite{vldb2014}, 2017~\cite{vldb2017}, 2018~\cite{vldb2018} \\
      KDD                 & 2015~\cite{kdd2015}, 2017~\cite{kdd2017}, 2018~\cite{kdd2018}    \\
      SOCC                & 2015~\cite{socc2015}, 2017~\cite{socc2017}, 2018~\cite{socc2018} \\
      OSDI / SOSP         & 2016~\cite{osdi2016}, 2017~\cite{sosp2017}, 2018~\cite{osdi2018} \\
      ICML                & 2016~\cite{icml2016}, 2017~\cite{icml2017}, 2018~\cite{icml2018} \\
      SC                  & 2016~\cite{sc2016}, 2017~\cite{sc2017}, 2018~\cite{sc2018}       \\
      \bottomrule
    \end{tabular}
\end{squishedtable}

For each of the 252 papers, we identified: (i)~the graph datasets used in
experiments; (ii)~the graph and machine learning computations that appeared in
the paper; and (iii)~the graph software used in the paper. In our survey, we
asked users questions about which graph and machine learning computations they
perform. The choices we provided in these questions came from the computations
we identified in these publications (see
Sections~\ref{subsec:graph-computations} and~\ref{subsec:ml-computations} and
Appendices~\ref{app:methodology-graph-computations}
and~\ref{app:methodology-ml-computations} for details).

\subsection{Review of Emails and Code
    Repositories}\label{sec:mailing-lists-review}

\noindent To answer some questions that participants' responses raised and to
identify more specific challenges users face than the ones we identified from
participants' responses, we reviewed emails in the mailing lists of the 22
software products between January and September of 2017. In addition, 20 of
these 22 software products had open source code repositories. We reviewed the
bug reports and feature requests (\emph{issues} henceforth) in these
repositories between January and September of 2017. We also reviewed the
repositories of 2 popular graph visualization tools: Gephi~\cite{gephi} and
Graphviz~\cite{graphviz}. For emails and issues before January 2017, we
performed a targeted keyword search to find more instances of the challenges we
identified in the January-September 2017 review.

In total, we reviewed over 6000 emails and issues. The overwhelming majority of
the emails and issues were routine engineering tasks, such as users asking how
to write a query or developers asking for integration with another software. The
number of emails and issues that were useful for identifying challenges were 299
in total. We review these challenges in Section~\ref{sec:email-challenges}.
Table~\ref{table:counts-emails-issues} in the appendix shows the exact number of
emails and issues we reviewed for each product, and the number of commits in
its code repository to give readers a sense of how active these repositories
are.

\subsection{Note on Methodology}%
\label{sec:note-on-methodology}
We end this section with two points about our methodology and a brief summary of the lessons we learned from performing a user study.

\vspace{0.2cm} \noindent \emph{Biases:} Performing a survey study brings up challenging
methodological questions, such as What is a principled way of recruiting
participants, picking a list of choices for graph queries, or reviewing academic
publications? Our guiding principle when addressing these questions was to be
as broad as possible and to avoid ad-hoc decisions. For example, the initial 22
graph software products we found were the products that had open source mailing
lists of a much longer list of all products we were aware of.

Similarly, when asking about the different graph queries and computations,
instead of an ad-host list of choices, we gathered a list from academic
publications. However, the choices we made inevitably introduced biases.

We acknowledge these biases when we present our findings in later
sections. In particular, the numbers we report should not be interpreted
statistically. Our goal was not to understand any statistical property, e.g.,
the average number of edges, about the graphs used in practice or users of
graphs. Despite these biases, we found overwhelming evidence for some of our
observations, which we believe give insights into
how graphs are used in practice.

\vspace{0.2cm} \noindent \emph{Abundance of Public Information:} Having direct access to actual
users from industry is a known challenge for researchers in academia. There is,
however, an abundance of public information in mailing lists, forums, vendor
websites, open source code repositories, social media, question-and-answer
websites, and elsewhere, which can be used to survey actual users and arrange
in-person interviews. In this paper, we essentially reviewed this public
information and used it to contact actual users. Our work is not the first but
the most extensive that we know of in terms of the public sources it
reviews. We believe our methodology can easily be repeated by researchers in
academia to study how other types of data or technology is used in practice.

\vspace{0.2cm} \noindent \emph{Lessons from the Survey Methodology:} We highlight three lessons from our experience of applying our methodology. 
\begin{squishedlist}

\item {\em Lesson 1:} Many users are willing to share information. Especially for our online survey, we did not expect to recruit 89 participants prior to sending the survey out. 

\item {\em Lesson 2:} Avoid making assumptions about participants' answers in the survey. For instance, as we discuss in Section~\ref{sec:graph-size}, we assumed few users would have edge graphs with more than 1 billion edges, so capped the choices of a question about graph sizes at 1 billion. This resulted in losing important information about how much larger the graphs are beyond 1B.

\item {\em Lesson 3:} Users have different and often non-technical languages than researchers to explain their technology. For instance, our interview with a biologist using an RDF engine heavily involved terms such as tissues, processes, angiogenesis, molecules and chemical reactions. This required spending considerable time during the interview on terminology, which put a hurdle on focusing on technical topics on graph processing. We learned to thoroughly study our inteviewee's products and application domains prior to an interview.
\end{squishedlist}

%% file: 03survey.tex
\section{Online Survey}
\label{sec:survey}

In this section, we describe the questions we asked in the survey and report the
responses of the participants. We also report the results of our review of
academic publications. Throughout the section, we highlight the results that we
found particularly interesting or surprising.

\subsection{Graph Datasets}\label{sec:graph-datasets}

\noindent In this section, we describe the properties of the graph datasets that
the participants work with.

\subsubsection{Real-World Entities Represented}\label{sec:real-world-entities}
\noindent We asked the participants about the real-world entities that their
graphs represent. We provided them with 4 choices and the participants could
select multiple of them.

\begin{squishedenumerate}

    \item \emph{Humans}: e.g., employees, customers, and their interactions.
    \item \emph{Non-Human Entities}: e.g., products, transactions, or web pages.
    \item \emph{RDF or Semantic Web}.
    \item \emph{Scientific}: e.g., chemical molecules or biological proteins.

\end{squishedenumerate}

\noindent For the participants who selected non-human entities, we followed up
with a short-answer question asking them to describe what these are.
Participants indicated 52 different kinds of non-human entities, which we group
into 7 broad categories.\footnote{Six entities that the participants mentioned
    did not fall under any of our 7 categories, which we list for completeness:
    call records, computers, cars, houses, time slots, and specialties.} We
indicate the acronyms we use in our tables for each category in parentheses:

\begin{squishedenumerate}

    \item \emph{Products} (NH-P): e.g., products, orders, and transactions.
    \item \emph{Business and Financial Data} (NH-B): e.g., business assets,
    funds, or bitcoin transfers.
    \item \emph{World Wide Web Data} (NH-W).
    \item \emph{Geographic Maps} (NH-G): e.g., roads, bicycle sharing stations,
    or scenic spots.
    \item \emph{Digital Data} (NH-D): e.g., files and folders or videos and
    captions.
    \item \emph{Infrastructure Networks} (NH-I): e.g., oil wells and pipes or
    wireless sensor networks.
    \item \emph{Knowledge and Textual Data} (NH-K): e.g., keywords, lexicon
    terms, words, and definitions.

\end{squishedenumerate}

\noindent Table~\ref{table:real-world-entities} shows the responses. In the
table, the number of academic publications that use each type of graph is listed
in the \emph{A} row. We highlight two interesting observations:

\begin{squishedtablestar}
    \centering
    \caption{Real-world entities represented by the participants' graphs and
        studied in publications. Legend for non-human entities:\ \emph{Products}
        (NH-P), \emph{Business and Financial Data} (NH-B), \emph{Web Data}
        (NH-W), \emph{Geographic Maps} (NH-G), \emph{Digital Data} (NH-D),
        \emph{Infrastructure Networks} (NH-I), \emph{Knowledge and Textual Data}
        (NH-K).}%
    \label{table:real-world-entities}
    \iftablesqueeze\vspace{-0.7em}\fi
    \begin{tabular}{lrrrrrrrrrrr}
      \toprule
      \textbf{Category} & \textbf{Human} & \textbf{RDF} & \textbf{Scientific} & \textbf{Non-Human} & \textbf{NH-P} & \textbf{NH-B} & \textbf{NH-W} & \textbf{NH-G} & \textbf{NH-D} & \textbf{NH-I} & \textbf{NH-K} \\ \midrule
      \textbf{Total}    & 45             & 23           & 15                  & 60                 & 13            & 11            & 4             & 7             & 5             & 9             & 11            \\
      \textbf{R}        & 18             & 11           & 9                   & 22                 & 1             & 6             & 2             & 4             & 1             & 7             & 6             \\
      \textbf{P}        & 27             & 12           & 6                   & 38                 & 12            & 5             & 2             & 3             & 4             & 2             & 5             \\
      \textbf{A}        & 165            & 20           & 45                  & 169                & 7             & 28            & 77            & 33            & 0             & 17            & 11            \\
      \bottomrule
    \end{tabular}
\end{squishedtablestar}

\begin{squishedtablestar}
    \centering
    \caption{The sizes of the participants' graphs.}
    \iftablesqueeze\vspace{-0.7em}\fi
    \begin{subtable}[t]{.32\textwidth}
        \centering \subcaption{Number of vertices.}%
        \label{table:no-of-vertices}
        \iftablesqueeze\vspace{-0.3em}\fi
        \begin{tabular}{lrrr}
          \toprule
          \multicolumn{1}{l}{\textbf{Vertices}} & \textbf{Total} & \textbf{R} & \textbf{P} \\ \midrule
          \textless{} 10K                       & 22             & 11         & 11         \\
          10K$-$100K                            & 22             & 9          & 13         \\
          100K$-$1M                             & 19             & 7          & 12         \\
          1M$-$10M                              & 17             & 6          & 11         \\
          10M$-$100M                            & 20             & 10         & 10         \\
          \textgreater{} 100M                   & 27             & 10         & 17         \\
          \bottomrule
        \end{tabular}
    \end{subtable}
    \begin{subtable}[t]{.32\textwidth}
        \centering \subcaption{Number of edges.}%
        \label{table:no-of-edges}
        \iftablesqueeze\vspace{-0.3em}\fi
        \begin{tabular}{lrrr}
          \toprule
          \textbf{Edges}                        & \textbf{Total} & \textbf{R} & \textbf{P} \\ \midrule
          \textless{} 10K                       & 23             & 11         & 12         \\
          10K$-$100K                            & 22             & 9          & 13         \\
          100K$-$1M                             & 13             & 3          & 10         \\
          1M$-$10M                              & 9              & 5          & 4          \\
          10M$-$100M                            & 21             & 8          & 13         \\
          100M$-$1B                             & 21             & 8          & 13         \\
          \textgreater{} 1B                     & 20             & 8          & 12         \\
          \bottomrule
        \end{tabular}
    \end{subtable}
    \begin{subtable}[t]{.32\textwidth}
        \centering \subcaption{Total uncompressed bytes.}%
        \label{table:uncompressed-bytes}
        \iftablesqueeze\vspace{-0.3em}\fi
        \begin{tabular}{lrrr}
          \toprule
          \textbf{Size}                         & \textbf{Total} & \textbf{R} & \textbf{P} \\ \midrule
          \textless{} 100MB                     & 23             & 12         & 11         \\
          100MB$-$1GB                           & 19             & 9          & 10         \\
          1GB$-$10GB                            & 25             & 9          & 16         \\
          10GB$-$100GB                          & 17             & 5          & 12         \\
          100GB$-$1TB                           & 20             & 8          & 12         \\
          \textgreater{} 1TB                    & 17             & 5          & 12         \\
          \bottomrule
        \end{tabular}
    \end{subtable}
\end{squishedtablestar}

\begin{squishedlist}

    \item \emph{Variety}: Real graphs capture a very wide variety of entities.
    Readers may be familiar with entities such as social connections,
    infrastructure networks, and geographic maps. However, many other entities
    in the participants' graphs may be less natural to think of as graphs. These
    include malware samples and their relationships, videos and captions, or
    scenic spots, among others. This lends credence to the clich\'{e} that
    graphs are everywhere.

    \item \emph{Product Graphs}: Products, orders, and transactions were the
    most popular non-human entities represented in practitioners' graphs,
    indicated by 12 practitioners. This contrasts with their relative
    unpopularity among researchers and academics. Only 1 researcher used product graphs and after digital data graphs, product graphs were the second least popular graphs used in academic papers. Such product-order-transaction data is traditionally the
    classic example of enterprise data that perfectly fits the relational data
    model. It is interesting that enterprises represent similar product data as
    graphs, possibly because they find value in analyzing connections in such
    data.

\end{squishedlist}

\noindent We also note that we expected scientific graphs to be used mainly by
researchers. Surprisingly, scientific graphs are prevalent among practitioners
as well.

\begin{squishedtable}
    \centering
    \caption{Sizes of organization using graphs with >1B edges.}%
    \label{table:large-graph-organizations}
    \iftablesqueeze\vspace{-0.7em}\fi
    \begin{tabular}{ccccc}
      \toprule
      \textbf{Size} & 1$-$10 & 10$-$100 & 100$-$1000 & \textgreater{} 10000 \\ \midrule
      \textbf{\#}   & 4      & 4        & 7          & 4                    \\
      \bottomrule
    \end{tabular}
\end{squishedtable}

\subsubsection{Size}\label{sec:graph-size}
\noindent We asked the participants the number of vertices, number of edges, and
total uncompressed size of their graphs. They could select multiple options.
Tables~\ref{table:no-of-vertices},~\ref{table:no-of-edges},
and~\ref{table:uncompressed-bytes} show the responses. As shown in the tables,
graphs of every size, from very small ones with less than 10K edges to very
large ones with more than 1B edges, are prevalent across both researchers and
practitioners. We make one interesting observation:

\begin{squishedtablestar}
    \centering
    \caption{The topology and stored data types of the participants' graphs.}%
    \label{table:graph-topology}
    \iftablesqueeze\vspace{-0.7em}\fi
    \begin{subtable}[t]{.28\textwidth}
        \centering \setlength\tabcolsep{0.25em} \subcaption{Directed vs.\
            Undirected}%
        \label{table:topology-directed}
        \iftablesqueeze\vspace{-0.3em}\fi
        \begin{tabular}{lrrr}
          \toprule
          \multicolumn{1}{l}{\textbf{Topology}}              & \textbf{Total}                        & \textbf{R} & \textbf{P}                                            \\ \midrule
          Only Directed                                      & 63                                    & 23         & 40                                                    \\
          Only Undirected                                    & 11                                    & 6          & 5                                                     \\
          Both                                               & 15                                    & 7          & 8                                                     \\
          \bottomrule
        \end{tabular}
    \end{subtable}
    \begin{subtable}[t]{.28\textwidth}
        \centering \setlength\tabcolsep{0.25em} \subcaption{Simple vs.\
            Multigraphs}%
        \label{table:topology-simple}
        \iftablesqueeze\vspace{-0.3em}\fi
        \begin{tabular}{lrrr}
          \toprule
          \multicolumn{1}{l}{\textbf{Topology}}              & \textbf{Total}                        & \textbf{R} & \textbf{P}                                            \\ \midrule
          Only Simple Graphs                                 & 26                                    & 9          & 17                                                    \\
          Only Multigraphs                                   & 50                                    & 20         & 30                                                    \\
          Both                                               & 13                                    & 7          & 6                                                     \\
          \bottomrule
        \end{tabular}
    \end{subtable}
    \begin{subtable}[t]{.4\textwidth}
        \centering \setlength\tabcolsep{0.45em}
        \caption{Data types stored on vertices and edges.}%
        \label{table:data-types}
        \iftablesqueeze\vspace{-0.3em}\fi
        \begin{tabular}{lrrrrrr}
          \toprule
          \multicolumn{1}{l}{\multirow{2}{*}{\textbf{Type}}} & \multicolumn{3}{c}{\textbf{Vertices}} & \multicolumn{3}{c}{\textbf{Edges}}                                 \\ \cmidrule{2-7}
          \multicolumn{1}{c}{}                               & \textbf{Total}                        & \textbf{R} & \textbf{P} & \textbf{Total} & \textbf{R} & \textbf{P} \\ \midrule
          String                                             & 79                                    & 31         & 48         & 66             & 24         & 42         \\
          Numeric                                            & 63                                    & 23         & 40         & 59             & 23         & 36         \\
          Date/Timestamp                                     & 56                                    & 19         & 37         & 49             & 18         & 31         \\
          Binary                                             & 15                                    & 8          & 7          & 8              & 4          & 4          \\
          \bottomrule
        \end{tabular}
    \end{subtable}
\end{squishedtablestar}

\begin{squishedlist}
    \item \emph{The Ubiquity of Very Large Graphs:} A significant number of
    participants work with very large graphs. Specifically, 20 participants (8
    researchers and 12 practitioners) indicated using graphs with more than a
    billion edges. Moreover, the 20 participants with graphs with more than one
    billion edges are from organizations with different scales, ranging from
    very small to very large, as shown in
    Table~\ref{table:large-graph-organizations}. This refutes the common
    assumption that only very large organizations---such as
    Google~\cite{Malewicz:2010:PSL:1807167.1807184},
    Facebook~\cite{Ching:2015:OTE:2824032.2824077}, and
    Twitter~\cite{DBLP:journals/pvldb/SharmaJBLL16} that have web and social
    network data---have very large graphs. Finally, we note that these large
    graphs represent a variety of entities, including social, scientific, RDF,
    product, and digital data,\footnote{Some participants selected multiple
        graph sizes and multiple entities, so we cannot perform a direct match
        of which graph size corresponds to which entity. The entities we list
        here are taken from the participants who selected a single graph size
        and entity, so we can directly match the size of the graph to the
        entity.} indicating that very large graphs appear in a wide range of
    domains.
\end{squishedlist}

\noindent One thing that is not clear from our survey is how much larger the
participants' graphs are beyond the maximum limits we inquired about (100
million vertices, 1 billion edges, and 1TB uncompressed data). In order to
answer this question, we categorized the graph sizes mentioned in the user
emails we reviewed that were beyond these sizes. Focusing on the number of
edges, we found 42 users with 1--10B-edge graphs, 17 with 10B--100B-edge graphs,
and 7 users processing graphs over 100B edges. Two participants also clarified
through an email exchange that their graphs contained 4B and 30B edges. As in
our survey results, these large graphs represented a wide range of entities,
such as product-order-transaction data, or entities from agriculture and
finance. Table~\ref{table:sizes-email} in the appendix shows the exact
distribution of sizes we identified. As we discuss in Section~\ref{sec:applications-interviews}, several of the applications described in our applications also contained graphs in the 10B-100B-edge and over 100B-edge scale.

\subsubsection{Other Questions on Graph Datasets}

\vspace{0.5em} \noindent \textbf{Topology}: We asked the participants whether
their graphs were: (i)~\emph{directed or undirected}; and (ii)~\emph{simple
    graphs or multigraphs}. We clarified that multigraphs are those with
possibly multiple edges between two vertices, while simple graphs do not allow
multiple edges between two vertices. Tables~\ref{table:topology-directed}
and~\ref{table:topology-simple} show the responses.

\vspace{0.5em} \noindent \textbf{Types of Data Stored on Vertices and Edges}: We
asked the participants whether they stored data on the vertices and edges of
their graphs. All participants except 3 indicated that they do. We asked the
types of data they store and gave them 4 choices: (i)~string; (ii)~numeric;
(iii)~date or timestamp; and (iv)~binary. Table~\ref{table:data-types} shows
participants' responses. Five participants also indicated storing JSON, lists,
and geographic coordinates using the \emph{Other} option.

\vspace{0.5em} \noindent \textbf{Dynamism}: We asked the participants how
frequently the vertices and edges of their graphs change, i.e., are added,
deleted, or updated. We provided 3 choices with the following explanations:
(i)~\emph{static}: there are no or very infrequent changes; (ii)~\emph{dynamic}:
there are frequent changes, and all changes are stored permanently; and
(iii)~\emph{streaming}: there are very frequent changes and the participants'
software discards some of the graph after some time.
Table~\ref{table:dynamism-frequency} shows the responses. Surprisingly 18
participants (9~researchers and 9~practitioners) indicated having streaming graphs.

\begin{squishedtable}
    \centering
    \caption{Frequency of changes.}%
    \label{table:dynamism-frequency}
    \iftablesqueeze\vspace{-0.7em}\fi
    \begin{tabular}{lrrr}
      \toprule
      \textbf{Frequency} & \textbf{Total} & \textbf{R} & \textbf{P} \\ \midrule
      Static             & 40             & 21         & 19         \\
      Dynamic            & 55             & 22         & 33         \\
      Streaming          & 18             & 9          & 9          \\
      \bottomrule
    \end{tabular}
\end{squishedtable}

\subsection{Computations}\label{sec:graph-computations}

\noindent In this section, we describe the computations that the participants
perform on their graphs.

\subsubsection{Graph Computations}\label{subsec:graph-computations}

\noindent Our goal in this question was to understand what types of graph
queries and computations, not including machine learning computations,
participants perform on their graphs. We asked a multiple choice question that
contained as choices a list of queries and computations followed by a short
answer question that asked for computations that may not have appeared in the
first question as a choice. In the multiple choice question, instead of asking
for a set of ad-hoc queries and computations, we selected a list of graph
queries and computations that appeared in the publications of the 6 conferences
we reviewed (recall Section~\ref{sec:academic-publications-review}), using our
best judgment to categorize similar computations under the same name. We
describe our detailed methodology in
Appendix~\ref{app:methodology-graph-computations}.

Table~\ref{table:graph-computations} shows the 13 choices we provided in the
multiple choice question, the responses we got, and the number of academic
publications that use or study each computation. As shown in the table, all of
the 13 computations are used by both researchers and practitioners. Except for
two computations, the popularity of these computations is similar among
participants' responses and academic publications. The exceptions are
neighborhood and reachability queries, which are respectively used by 51 and 27
participants, but studied respectively in 10 and 8 publications. Finding connected components
appears to be a very popular and fundamental graph computation---it is the most
popular graph computation overall and also among practitioners. We suspect it is
a common pre-processing or cleaning step, e.g., to remove singleton vertices,
across many tasks.

A total of 13 participants answered our follow-up short answer question on other
graph queries and computations they run. Example answers include queries to
create schemas and graphs, custom bio-informatics algorithms, and finding
$k$-cores in a weighted graph.

\begin{squishedtable}
    \centering \setlength{\tabcolsep}{0.4em}
    \caption{Graph computations performed by the participants and studied in
        publications.}%
    \label{table:graph-computations}
    \iftablesqueeze\vspace{-0.7em}\fi
    \begin{tabular}{p{5.5cm}rrrr}
      \toprule
      \multicolumn{1}{l}{\textbf{Computation}}                              & \textbf{Total} & \textbf{R} & \textbf{P} & \textbf{A} \\ \midrule
      Finding Connected Components                                          & 55             & 18         & 37         & 31         \\ \midrule
      Neighborhood Queries (e.g., finding 2-degree neighbors of a vertex)   & 51             & 19         & 32         & 9         \\ \midrule
      Finding Short / Shortest Paths                                        & 43             & 18         & 25         & 28         \\ \midrule
      Subgraph Matching (e.g., finding all diamond patterns, SPARQL)        & 33             & 14         & 19         & 52         \\ \midrule
      Ranking \& Centrality Scores (e.g., PageRank, Betweenness Centrality) & 32             & 17         & 15         & 45         \\ \midrule
      Aggregations (e.g., counting the number of triangles)                 & 30             & 10         & 20         & 24         \\ \midrule
      Reachability Queries (e.g.,checking if $u$ is reachable from $v$)     & 27             & 7          & 20         & 8          \\ \midrule
      Graph Partitioning                                                    & 25             & 13         & 12         & 12         \\ \midrule
      Node-similarity (e.g., SimRank)                                       & 18             & 7          & 11         & 11         \\ \midrule
      Finding Frequent or Densest Subgraphs                                 & 11             & 7          & 4          & 4          \\ \midrule
      Computing Minimum Spanning Tree                                       & 9              & 5          & 4          & 4          \\ \midrule
      Graph Coloring                                                        & 7              & 3          & 4          & 8          \\ \midrule
      Diameter Estimation                                                   & 5              & 2          & 3          & 2          \\
      \bottomrule
    \end{tabular}
\end{squishedtable}

\subsubsection{Machine Learning Computations}\label{subsec:ml-computations}

\noindent We next asked participants what kind of machine learning computations
they perform on their graphs. Similar to the previous question, these questions
were formulated to identify the machine learning computations that appeared in
the academic publications we reviewed. We describe our detailed methodology in
Appendix~\ref{app:methodology-ml-computations}. We asked the following 2
questions:

\begin{squishedlist}

    \item \emph{Which machine learning computations do you run on your graphs?}
    The choices were: clustering, classification, regression (linear or
    logistic), graphical model inference, collaborative filtering, stochastic
    gradient descent, and alternating least squares.

    \item \emph{Which problems that are commonly solved with machine learning do
        you solve using graphs?} The choices were: community detection,
    recommendation system, link prediction, and influence
    maximization.\footnote{In the publications, link prediction referred to
        problems that predict a missing edge in a graph or data on an existing
        edge. Influence maximization referred to finding influential vertices in
        a graph, e.g., those that can bring more vertices to the graph. We did
        not provide detailed explanations about the problems to the
        participants.}

\end{squishedlist}

\noindent Tables~\ref{table:ml-computations} and~\ref{table:ml-problems} show
the responses and the number of academic publications that use or study each
computation. It is clear that machine learning is used very widely in graph
processing. Specifically, 61 participants indicated that they either perform a
machine learning computation or solve a problem using machine learning on their
graphs. Clustering is the most popular computation performed, while community
detection is the most popular problem solved using machine learning. None of the
participants selected alternating least squares as a computation they perform.

\begin{squishedtable}
    \centering \setlength{\tabcolsep}{0.4em}
    \caption{Machine learning computations and problems performed by the
        participants and studied in publications.}%
    \label{table:machine-learning-computations}
    \vspace{-0.7em}
    \begin{subtable}[t]{\columnwidth}
        \centering \subcaption{Machine learning computations.}%
        \label{table:ml-computations}
        \iftablesqueeze\vspace{-0.3em}\fi
        \begin{tabular}{p{3.6cm}rrrr}
          \toprule
          \multicolumn{1}{l}{\textbf{Computation}} & \textbf{Total} & \textbf{R} & \textbf{P} & \textbf{A} \\ \midrule
          Clustering                               & 42             & 22         & 20         & 22         \\
          Classification                           & 28             & 10         & 18         & 34         \\
          Regression (Linear / Logistic)           & 11             & 5          & 6          & 2          \\
          Graphical Model Inference                & 10             & 5          & 5          & 5          \\
          Collaborative Filtering                  & 9              & 4          & 5          & 5          \\
          Stochastic Gradient Descent              & 4              & 2          & 2          & 9         \\
          Alternating Least Squares                & 0              & 0          & 0          & 1          \\ \bottomrule
        \end{tabular}
        \vspace{1em}
    \end{subtable}
    \begin{subtable}[t]{\columnwidth}
        \centering \subcaption{Problems solved by machine learning algorithms.}%
        \label{table:ml-problems}
        \iftablesqueeze\vspace{-0.3em}\fi
        \begin{tabular}{p{3.6cm}rrrr}
          \toprule
          \multicolumn{1}{l}{\textbf{Computation}} & \textbf{Total} & \textbf{R} & \textbf{P} & \textbf{A} \\ \midrule
          Community Detection                      & 31             & 15         & 16         & 15         \\
          Recommendation System                    & 26             & 10         & 16         & 5          \\
          Link Prediction                          & 25             & 10         & 15         & 11         \\
          Influence Maximization                   & 14             & 5          & 9          & 6          \\ \bottomrule
        \end{tabular}
    \end{subtable}
\end{squishedtable}

\subsubsection{Other Questions on
    Computations}\label{sec:streaming-computations}

\vspace{0.5em} \noindent \textbf{Streaming Computations}: We asked the
participants if they performed incremental or streaming computations on their
graphs:\ 32 participants (16 researchers and 16 practitioners) indicated that
they do. We followed up with a question asking them to describe the incremental
or streaming computations that they perform. A total of 4 participants indicated
computing graph or vertex-level statistics and aggregations; A total of 3
participants indicated incremental or streaming computation of the following
algorithms: approximate connected components, $k$-core, and hill climbing. For
completeness, we list the other computations participants mentioned: computing
node or community properties, calculating approximate answers to simple queries,
incremental materialization, incremental enhancement of the knowledge graph, and
scheduling.

We note that the 22 software products in Table~\ref{table:graph-software} have
limited or no support for incremental and streaming computations. We further note that we did not find any user in our further reviews of other sources or interviews that performed continuous computation on a very dynamic stream of edges or nodes.

\vspace{0.5em} \noindent \textbf{Traversals}: We asked the participants which
fundamental traversals, breadth-first search or depth-first search, they use in
their algorithms. Table~\ref{table:traversals} shows the responses. Participants
commonly use both kinds of traversals.

\begin{squishedtable}
    \centering
    \caption{Graph traversals performed by the participants.}%
    \label{table:traversals}
    \iftablesqueeze\vspace{-0.7em}\fi
    \begin{tabular}{p{3.6cm}rrr}
      \toprule
      \multicolumn{1}{l}{\textbf{Traversal}} & \textbf{Total} & \textbf{R} & \textbf{P} \\ \midrule
      Breadth-first-search or variant        & 19             & 5          & 14         \\
      Depth-first-search or variant          & 12             & 4          & 8          \\
      Both                                   & 22             & 8          & 14         \\
      Neither                                & 20             & 11         & 9          \\
      \bottomrule
    \end{tabular}
\end{squishedtable}

\subsection{Graph Software}\label{sec:graph-software}

\noindent We next review the properties of the different graph software that the
participants use.

\subsubsection{Software Types}\label{sec:software-non-query}

\noindent \textbf{Software for Querying and Performing Computations}: We asked
the participants which types of graph software they use to query and perform
computations on their graphs. The choices included 5 types of software from
Table~\ref{table:graph-software} as well as distributed data processing systems
(DDPSes), such as Apache Hadoop and Spark, relational database management
systems (RDBMSes), and linear algebra libraries and software, such as BLAS and
MATLAB.\@ Table~\ref{table:querying-software} shows the exact choices and
responses: 84 participants answered this question and each selected 2 or more
types of software. We highlight 3 interesting observations:

\begin{squishedtable}
    \centering \setlength{\tabcolsep}{0.4em}
    \caption{Software for graph queries and computations.}%
    \label{table:querying-software}
    \iftablesqueeze\vspace{-0.7em}\fi
    \begin{tabular}{p{5.5cm}rrrr}
      \toprule
      \multicolumn{1}{l}{\textbf{Software}}                           & \textbf{Total} & \textbf{R} & \textbf{P} & \textbf{A} \\ \midrule
      Graph Database System (e.g., Neo4j, OrientDB, TitanDB)          & 59             & 20         & 39         & 6          \\ \midrule
      Apache Hadoop, Spark, Pig, Hive                                 & 29             & 11         & 18         & 10         \\ \midrule
      Apache Tinkerpop (Gremlin)                                      & 23             & 9          & 14         & 1          \\ \midrule
      Relational Database Management System (e.g., MySQL, PostgreSQL) & 21             & 6          & 15         & 7          \\ \midrule
      RDF Engine (e.g., Jena, Virtuoso)                               & 16             & 8          & 8          & 12         \\ \midrule
      Distributed Graph Processing Systems (e.g., Giraph, GraphX)     & 14             & 8          & 6          & 36         \\ \midrule
      Linear Algebra Library / Software (e.g., MATLAB, Maple, BLAS)   & 8              & 6          & 2          & 6          \\ \midrule
      In-Memory Graph Processing Library (e.g., SNAP, GraphStream)    & 7              & 5          & 2          & 4          \\ \bottomrule
    \end{tabular}
\end{squishedtable}

\begin{squishedlist}

    \item \emph{Popularity of Graph Database Systems}: The most popular choice
    was graph database systems. We suspect this is partly due to their
    increasing popularity and partly due to the inherent bias in the
    participants we recruited---as explained in
    Section~\ref{sec:participant-recruitment}, more of them came from users of
    graph database systems. We did not ask the participants which specific graph
    database system they used.

    \item \emph{Popularity of RDBMSes}: 21 participants (6 researchers and 15
    practitioners) chose RDBMSes. We consider this number high given that we did
    not recruit participants from the mailing lists of any RDBMS.\@
    Interestingly, 16 of these 20 participants also indicated using graph
    database systems. From our survey, we cannot answer what the participants
    used RDBMSes for. It is possible that they use an RDBMS as the main
    transactional storage and a graph database system for graph-specific tasks
    such as traversals. This was the case in the applications described to us in our interviews (see Section~\ref{sec:applications-interviews}).

    \item \emph{Unpopularity of DGPSes}: Only 6 practitioners indicated using a
    DGPS, such as Giraph, GraphX, and Gelly. This contrasts with DGPSes'
    popularity among academics, where they are the most popular systems, studied
    by 36 publications. One can consider graph database systems as RDBMSes that
    are specialized for graphs and DGPSes as DDPSes that are specialized for
    graphs. In light of this analogy, we note that there is an opposite trend in
    the usage of these groups of systems. While more participants indicated
    using graph database systems than RDBMSes, significantly more participants
    indicated using DDPSes than DGPSes.

\end{squishedlist}

\vspace{0.5em} \noindent \textbf{Software for Non-Querying Tasks}: We asked the
participants which types of graph software, possibly an in-house one, they use
for tasks other than querying graphs. Table~\ref{table:non-querying-software}
shows the choices and the responses. We highlight one interesting observation:

\begin{squishedlist}

    \item \emph{Importance of Visualization:} Visualization software is, by a
    large margin, the most popular type of software participants use among the 5
    choices. This clearly shows that graph visualization is a very common and
    important task. As we discuss in Section~\ref{sec:graph-challenges},
    participants also indicated visualization as one of their most important
    challenges when processing graphs.

\end{squishedlist}

\begin{squishedtable}
    \centering
    \caption{Software used for non-querying tasks.}%
    \label{table:non-querying-software}
    \iftablesqueeze\vspace{-0.7em}\fi
    \begin{tabular}{p{3.3cm}rrrr}
      \toprule
      \textbf{Software}           & \textbf{Total} & \textbf{R} & \textbf{P} & \textbf{A} \\ \midrule
      Graph Visualization         & 55             & 22         & 33         & 15         \\
      Build / Extract / Transform & 14             & 8          & 6          & 0          \\
      Graph Cleaning              & 5              & 1          & 4          & 2          \\
      Synthetic Graph Generator   & 4              & 3          & 1          & 49         \\
      Specialized Debugger        & 2              & 0          & 2          & 0          \\ \bottomrule
    \end{tabular}
\end{squishedtable}

\begin{squishedtable}
    \centering
    \caption{Architectures of the software used by participants.}%
    \label{table:count-software-architecture}
    \iftablesqueeze\vspace{-0.7em}\fi
    \begin{tabular}{p{4.07cm}rrr}
      \toprule
      \multicolumn{1}{l}{\textbf{Architecture}} & \textbf{Total} & \textbf{R} & \textbf{P} \\ \midrule
      Single Machine Serial                     & 31             & 17         & 14         \\
      Single Machine Parallel                   & 35             & 21         & 14         \\
      Distributed                               & 45             & 17         & 28         \\
      \bottomrule
    \end{tabular}
\end{squishedtable}

\subsubsection{Other Questions on Software}

\noindent \textbf{Software Architectures}: We asked the participants the
architectures of the software products they use for processing graphs. The
choices were single machine serial, single machine parallel, and distributed.
Table~\ref{table:count-software-architecture} shows the responses. Distributed
products were the most popular choice and users' selections highly correlated
with the size of graphs they have. For example, 29 of the 45 participants that
selected distributed architecture had graphs over 100M edges.

\vspace{0.5em} \noindent \textbf{Data Storage in Multiple Formats}: We asked the
participants whether or not they store a single graph in multiple formats:\ 33
participants answered yes and the most popular multiple format combination was a
relational database format and a graph database format.
Appendix~\ref{app:dataformats} provides the detailed responses.

\subsection{Practical Challenges}\label{sec:graph-challenges}
\noindent In this section, we first discuss the challenges in graph processing
that the participants identified, followed by a discussion of the challenges
that we identified through our review of user emails and code repositories of
different types of graph technologies.

\subsubsection{Challenges Identified from
    Survey}\label{subsec:survey-challenges}

\noindent We asked the participants 2 questions about the challenges they face
when processing their graphs. First, we asked them to indicate their top 3
challenges out of 10 choices we provided. Table~\ref{table:graph-challenges}
shows the choices and the participants' responses. Second, we asked them to
state their biggest challenge in a short-answer question. Three major challenges
stand out unequivocally from the responses:

\begin{squishedtable}
    \centering
    \caption{Graph processing challenges faced by participants.}%
    \label{table:graph-challenges}
    \iftablesqueeze\vspace{-0.7em}\fi
    \begin{tabular}{p{5.4cm}rrr}
      \toprule
      \multicolumn{1}{l}{\textbf{Challenge}}                                                                   & \textbf{Total} & \textbf{R} & \textbf{P} \\ \midrule
      Scalability (i.e., software that can process larger graphs)                                              & 45             & 20         & 25         \\ \midrule
      Visualization                                                                                            & 39             & 17         & 22         \\ \midrule
      Query Languages / Programming APIs                                                                       & 39             & 18         & 21         \\ \midrule
      Faster graph or machine learning algorithms                                                              & 35             & 19         & 16         \\ \midrule
      Usability (i.e., easier to  configure and use)                                                   & 25             & 10         & 15         \\ \midrule
      Benchmarks                                                                                               & 22             & 12         & 10         \\ \midrule
      More general purpose graph software (e.g., that can process offline, online, and streaming computations) & 20             & 9          & 11         \\ \midrule
      Graph Cleaning                                                                                           & 17             & 7          & 10         \\ \midrule
      Debugging \& Testing                                                                                     & 10             & 2          & 8          \\ \bottomrule
    \end{tabular}
\end{squishedtable}

\begin{squishedlist}

    \item \emph{Scalability}: The ability to process large graphs is the most
    pressing challenge participants face. Scalability was the most popular
    choice in the first question for both researchers and practitioners.
    Moreover, it was the most popular answer in the second question where 13
    participants reiterated that scalability is their biggest challenge. The
    specific scalability challenges that the participants mentioned include
    inefficiencies in loading, updating, and performing computations, such as
    traversals, on large graphs.

    \item \emph{Visualization}: Perhaps more surprisingly, graph visualization
    emerges as one of the top 3 graph processing challenges, as indicated by 39
    participants in the first question and 1 participant in the short-answer
    question. This is consistent with the participants indicating visualization
    as the most popular non-query task they perform on their graphs, as
    discussed in Section~\ref{sec:software-non-query}.

    \item \emph{Query Languages and APIs}: Query languages and APIs present
    another common graph processing challenge, as indicated by 39 participants
    in the first question and 5 participants in the short-answer question. The
    specific challenges mentioned in the short-answer responses include
    expressibility of query languages, compliance with standards, and
    integration of APIs with existing systems. For instance, one participant
    found current graph query languages to have poor support for debugging
    queries and another participant indicated their difficulty in finding
    software that complies fully with SPARQL standards.

\end{squishedlist}

\subsubsection{Challenges Identified from Review}\label{sec:email-challenges}

\noindent To go beyond the survey and to understand more specific challenges
users face or new functionalities users want, we studied the user emails and
code repositories of different classes of software. Below, we categorize the
challenges we found on visualization in graph database systems, RDF engines,
DGPSes, and graph libraries, separately under {\em Visualization}. We also list
the challenges we found in graph database systems and RDF engines related to
query languages separately under \emph{Query Languages}. The exact counts of
emails and issues we found for each challenge is in
Table~\ref{table:email-challenges} in the appendix.

\vspace{0.5em} \noindent{{\bf Graph Database Systems and RDF Engines}}:

\begin{squishedlist}

    \item \emph{High-Degree Vertices}: Users want the ability to process very
    high-degree vertices in a special way. One common request is to skip finding
    paths that go over such vertices either for efficient querying or because
    users do not find such paths interesting.

    \item \emph{Hyperedges}: Hyperedges are edges between more than 2 vertices,
    e.g., a family relationship between three individuals. In graph database
    systems and RDF engines, there is no native-way to represent hyperedges. The
    user discussions include suggestions to simulate hyperedges, such as having
    a ``hyperedge vertex'' and linking the vertices in the hyperedge to this
    mock vertex.

    \item \emph{Versioning and Historical Analysis}: Users want the ability to
    store the history of the changes made to the vertices and edges and query
    over the different versions of the graph. These requests are made in systems
    that do not support versioning and the discussions are on how to add
    versioning support at the application layer.

    \item \emph{Schema and Constraints}: Users want the ability to define
    schemas over their graphs, analogous to DTD and XSD schemas for XML
    data~\cite{dtd-xsd}, usually as a means to define constraints over their
    data. Examples include enforcing that the graph is acyclic or that some
    vertices always have a certain property.

    \item \emph{Triggers}: Users ask for trigger-like capabilities in their
    graph database systems. Examples include automatically adding a particular
    property to vertices during insertion or creating a backup of a vertex or an
    edge in the filesystem during updates. We note that some systems do support
    limited trigger functionality, such as OrientDB's \emph{hooks} or Neo4j's
    \emph{TransactionEventHandler} API.\@

\end{squishedlist}

\vspace{0.5em} \noindent{{\bf Graph Visualization Software}}:

\begin{squishedlist}

    \item \emph{Customizability:} One common challenge is to have the ability to
    customize the layout and design of the rendered graph, such as the shape or
    color of the vertices and edges.

    \item \emph{Layout}: Another common challenge is drawing graphs with certain
    structures on the screen according to a specific layout users had in mind.
    The most common example is drawing \emph{hierarchical} graphs, i.e., those
    in which some vertices are drawn on top of other vertices in an
    organizational hierarchy. Other examples include the drawing of star graphs,
    planar graphs, or a specialized tree layout, such as a \emph{phylogenetic
        tree}~\cite{letunic2006interactive}.

    \item \emph{Dynamic Graph Visualization:} Several users want support for or
    have challenges in animating the additions, deletions, and updates in a
    dynamic graph that is changing over time.

\end{squishedlist}

\noindent Users also have challenges in rendering large graphs with thousands or
even millions of vertices and edges.

\vspace{0.5em} \noindent{{\bf Query Languages}}: One of the most popular
discussions in user emails of graph database systems and RDF engines was writing
different queries in the query language of the software. In almost every case,
there was a way of satisfying the users' needs. Below we list 2 such types of
queries that could be interesting to researchers.

\begin{squishedlist}

    \item \emph{Subqueries}: Many users have challenges in the expression or
    performance of subqueries, i.e., using a query as part of another query. The
    challenges vary across different systems. Some users want the ability to
    embed SQL as a subquery in SPARQL.\@ Other users want the results of a
    subquery to be a graph that can further be queried,\footnote{This feature is
        called \emph{composition} and is supported in SPARQL but not in the
        languages of some graph database systems.} or to use a subquery as a
    predicate in another query.

    \item \emph{Querying across Multiple Graphs}: A common request in graph
    database systems and RDF engines is to construct queries that span multiple
    graphs, such as using the results of a traversal in one graph to start
    traversals in another. This is analogous to querying over multiple tables by
    joins in RDBMSes.\footnote{This functionality is supported in RDF engines
        but not supported in some graph database systems.}

\end{squishedlist}

\noindent Profiling and debugging slow queries and using indices correctly to
speed up queries are other common topics among users.

\vspace{0.5em} \noindent{{\bf DGPSes and Graph Libraries}}:

\begin{squishedlist}
    \item \emph{Off-the-Shelf Algorithms:} The most common request we found in
    DGPSes and graph libraries is the addition of a new algorithm that users
    could use off-the-shelf. All of these software products provide lower-level
    programming APIs using which users can compose graph algorithms. A small
    number of users want enhancements to these APIs as well. From our review, it
    appears that users of these software products find more value in directly
    using an already implemented algorithm than implementing the algorithms
    themselves.
    \item \emph{Graph Generators:} All of the DGPSes and graph libraries in our
    list have modules to generate synthetic graphs. Our review revealed that
    users find these graph generators useful, e.g., for testing algorithms. A
    common request was the ability to generate different kinds of synthetic
    graphs, such as $k$-regular graphs or random directed power-law graphs.
    \item \emph{GPU Support:} Several users, both in DGPSes and graph libraries,
    want support for running graph algorithms on GPUs.
\end{squishedlist}

\noindent In every DGPS we reviewed, a common challenge is users' computations
running out of cluster memory or having problems when using disk. We also note
that except for Gelly, every DGPS and every graph library either have a
visualization component or users have requests to add one, showing the
importance of visualization across users of a range of different graph
technologies.

\subsection{Workload Breakdown}\label{sec:graph-management}

\noindent We asked the participants how many hours per week they spend on 6
graph processing tasks and provided them with 3 choices: (i)~less than 5 hours;
(ii)~5 to 10 hours; and (iii)~more than 10 hours.
Table~\ref{table:graph-tasks-time} shows the choices and ranks the tasks in
terms of the number of participants that selected more than 10 hours first, then
5 to 10 hours, and then less than 5 hours. According to this ranking, the
participants spend the most time in analytics and testing and the least time on
ETL and cleaning.

\begin{squishedtable}
    \centering \setlength{\tabcolsep}{0.4em}
    \caption{Time spent by the participants on different tasks.}%
    \label{table:graph-tasks-time}
    \iftablesqueeze\vspace{-0.7em}\fi
    \begin{tabular}{lccc}
      \toprule
      \multicolumn{1}{l}{\textbf{Task}} & \textbf{0$-$5 hours} & \textbf{5$-$10 hours} & \textbf{\textgreater{} 10 hours} \\ \midrule
      Analytics                           & 30                   & 18                    & 23                             \\
      Testing                             & 40                   & 12                    & 20                             \\
      Debugging                           & 37                   & 18                    & 15                             \\
      Maintenance                         & 46                   & 14                    & 13                             \\
      ETL                                 & 44                   & 14                    & 10                             \\
      Cleaning                            & 52                   & 10                    & 6                              \\
      \bottomrule
    \end{tabular}
\end{squishedtable}

%% file: 04applicationswhitepapers.tex
\section{Applications from Whitepapers}\label{sec:applications-whitepapers}

\subsection{Methodology}%
\label{subsec:whitepaper-methodology}
In order to understand the popular application areas and fields
using graph software, we surveyed the whitepapers of software vendors. Whitepapers are documents that software vendors provide, often for marketing purposes, to give information about the use cases of their products. In our case, we consider whitepapers to be any document found on a software vendor's official website categorized as a whitepaper, a use case, a case study, or a scenario. From the initial software products in Table~\ref{table:graph-software}, only four graph database systems, specifically ArangoDB, Neo4j, OrientDB and Sparksee had whitepapers. To extend our review, we add the whitepapers of four RDF engines that were not in our initial list: AllegroGraph\cite{allegrograph}, AnzoGraph\cite{anzograph}, GraphDB by Ontotext\cite{ontotext}, and Stardog\cite{stardog}. We note that we only found whitepapers for graph database systems and RDF engines.

For each whitepaper, we selected the ones that describe an application using the product, e.g., music recommendation or money laundering detection. We omitted whitepapers that did not describe a specific application. For example, we omitted whitepapers specific to the software architecture of a product. In the end, we reviewed 89 whitepapers.

\subsection{Applications}
We labeled each whitepaper with a high-level application category and the field of industry of the customer in the case study.
Table~\ref{table:whitepapers} shows the different applications, the fields of industry in which the application was covered, and the number of whitepapers from graph databases and RDF systems that discussed the project. We found a total of 12 applications described in the whitepapers of graph databases and 5 applications in the whitepapers of RDF systems. As seen in the table, there is an overlap of the applications across both types of systems.

The three most popular applications were as follows:

\begin{squishedtablestar}
    \centering \setlength{\tabcolsep}{0.3em}
    \caption{Application areas and example uses of graphs in various fields
        described in graph software whitepapers.}%
    \label{table:whitepapers}
    \iftablesqueeze\vspace{-0.7em}\fi
    \begin{tabular}{p{2.9cm}p{4.1cm}p{5.8cm}rrr}
      \toprule
      \textbf{Application} & \textbf{Example} & \textbf{Fields} & \textbf{GDB} & \textbf{RDF} & \textbf{Total} \\ \midrule

      Data Integration
      	& Building an ontology by integrating multiple heterogeneous biomedical data sources
      	& Aerospace, Art \& Culture \& Heritage, Education, Entertainment, Finance, Food \& Cooking, Government, Health \& Life Sciences, Intelligence \& Law Enforcement, IT, Journalism, Marketing, Retail, Social Media, Toys \& Figurines

      	& 23 & 21 & 44
      \\ \midrule

      Personalization \& Recommendation
      	& Recommending products on an e-commerce platform
      	& Entertainment, Finance, Health \& Life Sciences, Hospitatility \& Travel, IT, Manufacturing, Marketing, Media, Music, Retail, Social Media, Telecommunication

      	&19 & 5 & 24
      \\ \midrule

      Fraud \& Threat Detection
      	& Detect cybercrime by searching for anomalous patterns
      	& Finance, Government, Insurance, Media, Retail

      	& 9 & 1 & 10
      \\ \midrule

      Risk Analysis \& Compliance
      	& Risk reporting by banks to comply with government regulations.
      	& Finance, Health \& Life Sciences, IT, Supply Chain \& Logistics

      	& 2 & 3 & 5
      \\ \midrule
      
      Identity \& Access Management
      & Monitor direct and indirect owners of businesses for financial analysis
      & Insurance, IT, Telecommunication

      & 4 & 0 & 4
      \\ \midrule

      Infrastructure Management \& Monitoring
      	& Manage cascading failures by tracking server interdependencies
      	& Intelligence \& Law Enforcement, IT

      	& 3 & 0 & 3
      \\ \midrule

Delivery \& Logistics
      	& Routing and tracking delivery parcels
      	& Retail, Supply Chain \& Logistics

      	& 2 & 0 & 2
      \\ \midrule

      Social Network Analysis
      	& Find the most viral users with maximum reach to other users
      	& Social Media

      	& 2 & 0 & 2
      \\ \midrule

      Other Applications
      & Natural language question answering, Call graph analysis, Code analysis, Drug discovery, Traffic route recommendation & IT, Telecommunication, Traffic Management & 3 & 2 & 5 \\
      \bottomrule

    \end{tabular}
\end{squishedtablestar}

\begin{squishedlist}
	\item \emph{Data Integration}: 44 whitepapers discussed primarily some data integration task that constructed a central, highly heterogenous graph from multiple sources. Data integration was also referred by some whitepapers as \emph{master data management} or \emph{knowledge graph creation}.  This category does not contain the whitepapers that described primarily another main business application but performed data integration as an initial step.

	Data integration whitepapers briefly mentioned a variety of other applications that would be supported by the integrated data, such as enterprise search. Many of these 44 whitepapers, as well as many whitepapers that discussed a data integration initial step, emphasized that customers found data integration easier in the semi-structured graph models than structured relational tables.

\item \emph{Personalization \& Recommendations}: The second most popular application was the use of a graph-based application data to personalize user interactions and provide better recommendations for the customers of a business. For example, one whitepaper described an e-commerce website that created a graph representation of the behavior of online shoppers and the interactions between customers and products to help make new product recommendations~\cite{arrangodb-aboutyou}. Another example was a personalized course curriculum service based on a hierarchical course topic relationships, represented as a graph, and the individual progress of each student~\cite{agensgraph-education}.

Many whitepapers avoided technical terms, but the applications described seemed to read the neighborhoods of users, represented as nodes in the underlying graph, to retrieve useful signals to make a recommendation.

\item \emph{Fraud \& Threat Detection}: The third most popular application was the detection of fraud and threats in various businesses. For example, one whitepaper described the use of graphs to detect financial fraud in banks by looking for rings in the graph formed after linking bank accounts, personal details, and financial transactions~\cite{neo4j-fraud}. Another application detected and prevented cyber crimes by monitoring for anomalous patterns in network traffic~\cite{agensgraph-cybercrime} represented as a graph.

    As a key benefit of using graphs, whitepapers highlighted that, compared to their equivalent SQL formulations, fraud patterns were easier to express as subgraph queries. Several whitepapers also mentioned that relational systems were not efficient enough to support these queries.

\end{squishedlist}

%% file: 05applicationsinterviews.tex
\section{Applications from Interviews}%
\label{sec:applications-interviews}

\subsection{Methodology}
\label{subsec:interview-methodology}
Whitepapers give an overview of the important applications using different
software but often contain very high level and non-technical
marketing language. To understand some of the applications using graphs in more
depth, we invited the participants of our
online survey for an in-person interview. 33 participants had provided us with their email addresses and 4 of them agreed.
To extend our interviews, we reached out to several of our contacts in major
software companies and graph vendors. We did 4 additional in-person interviews;
2 developers and 2 users of  graph processing software in major enterprises.

The occupations of our interviewees were as follows:
\begin{squishedlist}
    \item Two IT consultants to several large enterprises on graph technologies.
    \item A developer of graph processing systems at Alibaba.
    \item A developer of graph processing systems at Siemens.

    \item A principal scientist at Amazon working on knowledge graphs.
    \item Engineers from a contact management company called FullContact~\cite{full-contact}, an electric utility company called State Grid~\cite{state-grid}, and a startup called OpenBEL~\cite{openbel} that develops data publishing tools for biologists.
\end{squishedlist}

\noindent

 We lead the interviews with an open-ended question asking the interviewee to walk us through a concrete business application that uses graph data. The developers explained the applications of their customers. We asked questions about the details of the graph data, the graph computations they run, and the graph processing software they use in their applications. In addition, we asked three extra questions to each interviewee: (i)~Where do you use graph visualization? (ii)~Do you do streaming computations on your graphs? (iii)~Do you have machine learning computations that use your graphs? 

\subsection{Overall Observations}
We make four observations:
\begin{squishedlist}
\item None of the applications that used graphs representing transactional business data used a graph database or an RDF store as the main system of record. In each case, a relational system was the main system of record and the transactional data was replicated to a graph software for the application to use. This gives a sense of where graph databases and RDF systems are in the IT ecosystem of our interviewees' enterprises. The developer from Alibaba mentioned data replication as an important challenge for their internal customers.
\item Interviewees mentioned visualizing graphs in data exploration, debugging, query formulation, and as a presentation tool within the enterprise, for example to show a manager the benefits of modeling an application data as a graph.
\item Our interviewees were not aware of any continuous streaming computation performed on their graphs. Several interviewees mentioned processing highly dynamic graphs and buffering a window of several hours or days of these graphs. However, the computations in those applications were batch computations. For example, in one case, 3 days of business data would be copied over into a graph software to search for subgraph patterns. 

\item The machine learning applications interviewees discussed used graphs to extract features about nodes in a graph that were representing a business entity, such as a product or a customer. The feature extractions involved aggregating data from several hop neighborhoods of nodes and in one case through a recursive path query. These features would be used in vector representations of nodes and used by a machine learning application to do a prediction. We describe such a use case in Section~\ref{sec:siemens-automation}.
\end{squishedlist}

\noindent We next describe some of the applications from our interviews in detail. We discuss several other applications in Appendix~\ref{appendix:other-interviews}\@. Overall there were similarities between the applications described by our interviewees and those from the whitepapers but we also discovered some new applications. We cover one such new application called \emph{contingency analysis} in Section~\ref{sec:contingency-analysis}.
Some interviewees modified or omitted detailed information about their applications, data sets, queries, software, or challenges. For example, they modified what the vertices and edges actually are or gave the approximate sizes of their graphs. We report the applications as described by the interviewees.

\subsection{Recommendations}
\input{05a_alibaba_product_recommendation_ecommerce}

\input{05b_seimens_product_construction}

\input{05f_fraud_detection}

\input{05c_alibaba_amazon_question_answering}

\input{05d_contingency_analysis}

%% file: 05a_alibaba_product_recommendation_ecommerce.tex
\subsubsection{Keyword Recommendations on Alibaba's E-commerce Website}

When customers enter keywords to the search text box on Alibaba's e-commerce website, several keywords that are related to the search are recommended. These recommended keywords often aim to increase the diversity of products the user sees on the site. Internally, these recommendations are made by an application that uses a very large knowledge graph and performs parallel traversals that find and rank paths in this knowledge graph. There are two interesting aspects of this application:
\begin{squishedenumerate}
\item Strict latency: The recommendation needs to be done in several milliseconds. None of the other applications we encountered during our interviews required such strict latencies for the computations they had to perform. 
\item  Size and generation of the graph:  The underlying graph is primarily automatically generated from other data sources and was one of the largest graphs we encountered during our interviews. 
\end{squishedenumerate}

We describe the graph, the computation performed on the graph, and the software that stores the graph and performs the computation.

\customheader{Knowledge Graph:}\footnote{Note that the use of term ``knowledge graph'' vs other terms such as ``property graph'' or simply ``graph'' is slightly arbitrary. We found our interviewees referring to any data stored in RDF stores as knowledge graphs. We also found that interviewees referred to graphs
 that represent abstract things, e.g., keyword topics or concepts, also as knowledge graphs even if they were not stored in an RDF system.}
  The graph contains three types of nodes:
 \begin{squishedenumerate}
 \item {\em Products:} A subset, approximately 10 million, of products sold on Alibaba. 
 \item {\em Product Categories:} Includes categories such as ``shoes'', ``winter jackets'', ``TVs'', or ``electronics''. 

 There are approximately 10 thousand of categories.
 \item {\em Concepts:} This is an umbrella term to refer to a very large number of concrete and abstract real-world entities. Examples include ``football'', ``sports'', ``China'', ``young male'' or ``born in 80s''.  A small part of the concepts are manually curated inside Alibaba and some are obtained from the Chinese edition of Wikipedia. However, majority of them are previous search keywords that users have used. These are extracted from the search logs. As we describe momentarily, these are also the keywords that the application recommends to users. There are approximately 100 million of these.
\end{squishedenumerate}
There are two main edge types:
 \begin{squishedenumerate}
\item {\em Product-category edges:} In most cases each product belongs to exactly one category. So there are approximately 10 million of these.
\item {\em Concept-product and concept-category edges:}  There are edges between concepts and products and between concepts and categories, indicating a relation between a concept and a product or a category. The edges are automatically generated through several techniques, such as an analysis of the logs that contain keywords used by users, their clicks, and purchases or using natural language processing on reviews. To each generated edge, a weight is assigned to indicate the strength of the connection between the concept and the product or category. There are over 100 billion of these edges. 
\end{squishedenumerate}

\customheader{Recommendation Computation}: The recommendation computation happens as follows: Each user is tagged with a subset of the concepts indicating their known properties, such as ``male'' vs ``female'', ``IT professional'' vs ``accountant'' or ``born in 80s''. There are several dozen such tagged concepts. From each tagged concept a roughly 4-hop breadth-first search traversal is performed to find new concepts.
Each path from a tagged concept to each newly found concept $c$ is given a weight, based on the weights of the edges on the path, and aggregated to give a relevance weight to $c$. All new concepts are finally ranked and a subset of them are returned to the users. This entire computation has to happen in 4ms.

\customheader{Software:} The graph is stored in an in-house distributed graph database. The database stores the structure and properties on the nodes and edges in a distributed key-value store. So, both the neighbors and weights are stored as key value pairs and the 4-hop neighborhoods of nodes often need to be fetched from different machines.
Part of the graph is kept in memory and part is kept in SSDs.  The database supports the Gremlin language~\cite{gremlin} and the property graph model. The traversals are written in the Gremlin traversal API and resulting paths are aggregated in custom code written outside of Gremlin.

%% file: 05b_seimens_product_construction.tex
\subsubsection{\em Configuration Recommendation: Siemens' Automation Systems}
\label{sec:siemens-automation}

The context of our next application, also described briefly here~\cite{kg-siemens}, is the configuration of industrial automation systems, a project by Siemens.
Such systems are comprised of a combination of mechanical, hydraulic, and electric devices with complex constraints between individual component and a multitude of choices---for example, the number of input and output ports, line voltage, budget vs.\ premium options, etc.
A user, which might be an end customer or a sales representative, incrementally builds a plan that fulfills all the required functionalities, selecting from a product catalog.
While some configuration information is explicitly captured in terms of product features, there is also a large amount of tacit or implicit knowledge.
One solution that Siemens has been exploring is the use of recommendation techniques to aid in the selection of components.

The approach combines product information as well as past user behavior, modeled as a knowledge graph and stored in an RDF system.
Product information comes from a domain ontology and captures semantic relations such as ``has line voltage'', ``is of type'', and ``contains'', as well as product hierarchies, e.g., ``S7-1500'' is a subtype of ``S7'', which in turn is a specialization of a ``Control System''.
Information of past behavior comes from historical solutions, i.e., automation solutions that have been previously configured.

The novelty of this project comes from the combination of these two sources of evidence.
When asked as to why a knowledge graph as opposed to relational tables for storing and integrating these heterogeneous sources, our interviewee responded with two main reasons:\ first, the flexibility of the data model, and second, a knowledge graph is closer to how users conceptualize the data.
Both of these points were echoed by the white papers and other interviewees.

Although the graph database is an important component of the overall solution, its role is little more than a repository of features.
The actual recommendation algorithm is based on tensor factorization:\ the rows and columns correspond to entities (tens of thousands), while each slice corresponds to one relation, e.g., ``contains''.
Given the entities in a partial solution, the system's task is to recommend the most likely item to complete the solution (using previously-configured solutions as ground truth).
That is, the tensor is materialized from the graph database and used to train a model (written in TensorFlow in this case).
Thus, while this is perhaps an example of a machine-learning application on graph databases in a technical sense, the integration is rather shallow.

%% file: 05f_fraud_detection.tex
\subsection{Fraud and Threat Detection}
Four applications described in our interview with Alibaba and one application described by one consultant to a large financial institution were related to fraud and threat detection. There were two commonalities between these applications:
\begin{squishedenumerate}
\item {\em Searching a subgraph pattern:} Each application was based on finding some subgraph pattern, e.g., a bipartite, star, or cycle, in a very large transaction graph.
\item {\em Use of graph visualization:} In each case the detected pattern was merely a signal of a potential fraud that triggered an inspection by other systems or a  human for further investigation. Manual human investigations involved using a graph visualization software and exploring the neighborhoods of the emerged pattern. The consultant also mentioned using visualization for discovering the pattern to search for in the first place.
\end{squishedenumerate}

\noindent We briefly review some of these applications and when possible discuss the patterns searched.

\begin{figure}[t]
	\centering
	\captionsetup{justification=centering}
	\begin{subfigure}[b]{0.23\textwidth}
		
		\begin{subfigure}[b]{0.23\textwidth}
			\centering
			\begin{tikzpicture}[scale=0.7, transform shape,->,>=stealth', shorten >=1pt, auto, thick, node distance=2cm, main node/.style={circle,draw,font=\sffamily\Large\bfseries}]
			\node[main node] (1) {$P_1$};
			\node[main node] (2) [below right of=1] {$A_1$};
			\node[main node] (3) [above right of=1] {$A_3$};
			\node[main node] (4) [above right of=2] {$A_2$};
			\path[every node/.style={font=\sffamily\small}]
			(1) edge node [xshift=-28pt][yshift=-12pt] {$Owns$} (2) edge node  {$Owns$} (3)
			(2) edge node [xshift=70pt][yshift=-13.4pt] {$transfer\, \$100.00$} (4)
			(4)	edge node [xshift=70pt][yshift=12.2pt] {$purchase\, \$95.00$} (3) ;
			\end{tikzpicture}
			\vspace{0.2pt}
		\end{subfigure}

		\begin{subfigure}[b]{0.23\textwidth}
			\centering
			\begin{tikzpicture}[scale=0.7, transform shape,->,>=stealth', shorten >=1pt, auto, thick, node distance=2cm, main node/.style={circle,draw,font=\sffamily\Large\bfseries}]
			\node[main node] (1) {$P_1$};
			\node[main node] (2) [right of=1] {$A_3$};
			\node[main node] (3) [below right of=2] {$A_2$};
			\node[main node] (4) [below left of=3] {$A_4$};
			\node[main node] (5) [left of=4] {$P_2$};
			\path[every node/.style={font=\sffamily\small}]
			(1) edge node [xshift=-1pt][yshift=1pt] {$Owns$} (2) edge [<->] node [xshift=0pt] {$Friends$} (5)
			(3)	edge node [xshift=70pt][yshift=12.2pt] {$purchase\, \$95.00$} (2)
			(4) edge node [xshift=70pt][yshift=-13.4pt] {$transfer\, \$100.00$} (3)
			(5) edge node [xshift=-1pt][yshift=1pt] {$Owns$} (4) ;
			\end{tikzpicture}
		\end{subfigure}
		\caption{Cycle patterns.}%
		\label{fig:cyclic-pattern}
	\end{subfigure}
	\begin{subfigure}[b]{0.23\textwidth}
		\centering
		\begin{tikzpicture}[scale=0.7, transform shape,->,>=stealth', shorten >=1pt, auto,node distance=2cm, thick, main node/.style={circle,draw,font=\sffamily\Large\bfseries}]
		\node[main node] (0) {$B$};
		\node[main node] (1) [right of=0] {$P_2$};
		\node[main node] (2) [above of=1] {$P_1$};
		\node[main node] (3) [below of=1] {$P_k$};
		\node[main node] (4) [above right of=2] {$U_1$};
		\node[main node] (5) [below of=4] {$U_2$};
		\node[main node] (6) [below right of=3] {$U_n$};
		\path[every node/.style={font=\sffamily\small}]
		(0) edge (1) edge node {$Sells$} (2) edge (3)
		(4) edge (1) edge node [xshift=-34pt][yshift=12pt] {$Bought$} (2) edge (3)
		(5) edge (1) edge (3)
		(6) edge (1) edge (3) ;
		\path[-,every node/.style={font=\sffamily\small}]
		(1) edge[-,dotted] (3)
		(5) edge[-,dotted] (6) ;
		\end{tikzpicture}
		\caption{Bipartite transaction pattern.}%
		\label{fig:bipartite-transaction-pattern}
	\end{subfigure}
	\caption{Patterns for detecting fake transactions.}%
	\label{fig:patterns}
\end{figure}

\subsubsection{Fake Transactions on Alibaba.com} This application detects fake transactions initiated by businesses that sell products on Alibaba's e-commerce platform to increase their ranking on the platform.  There are two broad patterns the application searches for: (i) cycle patterns shown in Figure~\ref{fig:cyclic-pattern}; and (ii) a near-complete bipartite clique shown in Figure~\ref{fig:bipartite-transaction-pattern}. These patterns are detected by different applications on different graphs. We describe the patterns, the input graphs on which the patterns are searched, the software that searches the patterns, and several challenges our interviewee mentioned.

\subsubsection*{Cycle Patterns}

\customheader{Pattern:} The top pattern in Figure~\ref{fig:cyclic-pattern} is searching for an evidence that there is a business owner \texttt{P1} who is paying a fake buyer \texttt{P2} to buy products from \texttt{P1}. In particular, the pattern searches for transaction where \texttt{P1} transfers some amount of money from her account \texttt{A1} to an account \texttt{A2}, belonging to the fake buyer, which transfers a similar amount of money back to an account \texttt{A3} that is also owned by \texttt{P1}.  In a slightly more advanced version of the pattern, also discussed in a recent publication~\cite{qiu:cycle}, a friend of \texttt{P1} sends some amount of money to the fake buyer to initiate the fake purchase. Our interviewee noted that these cycle patterns are simplified versions of multiple other fraud patterns searched by other internal and external customers of Alibaba on other transaction graphs. Detecting such cyclic patterns in fraud-related applications also appeared in 7 use cases in the whitepapers.

\customheader{Input Graph:} These cyclic patterns are continuously searched
for in a graph that contains data about the financial accounts of the businesses on Alibaba, as well some social connection information, e.g., contact information of Alibaba customers or other available social information. 

\customheader{Software:} The current software stack is a bit complex, but briefly involves the following steps: (i) a stream of financial transaction edges are buffered for a period of time, roughly 10 seconds; (ii) the necessary neighborhood of those edges are extracted from a distributed in-house graph database; and (iii) the pattern is searched on the extracted graph in an single-node special solution. The extracted graph is several millions of edges and vertices. The applications latency is roughly 30 seconds.

\subsubsection*{Bipartite Patterns}

\customheader{Pattern:} Detecting fraud through cycle patterns is difficult because often money transfers do not go through Alibaba's systems. A more effective way is to find near-complete (and not necessarily fully complete) bipartite graph of products and customers on a graph extracted from the actual transactions on the Alibaba platform.  

Such patterns are signals that a large number of fake customers buy the same set of products, say owned by the same business.
This activity is similar to click farming~\cite{click-farm} to increase ad revenue of websites. Part of Alibaba's fraud detection system searches for multiple large instances of the pattern, where the pattern can contain hundreds of products and customers.

\customheader{Input Graph:} In a simplified form, the application runs on a graph that contains businesses, products, and customers as separate nodes and \texttt{sells} edges between businesses and products, \texttt{purchased} edges between products and customers. These patterns are very complex and detecting them on a large window of purchase transactions is very challenging, so the application limits the window to several days of transactions. This generates an input graph with a few hundred million nodes, and several billion edges.

\customheader{Software:} The application runs offline and uses a single-node custom-built in-memory graph processing software. 

\customheader{Challenges:} One challenge is to detect the nodes that are part of a pattern without enumerating the instances of these patterns. Enumeration of patterns that have a high-degree of symmetry is expensive because across two matches of the pattern, there can be a large overlap of the nodes. As a simple example, consider searching for a ($10$, $10$) complete bipartite pattern and the input graph contains a ($20$, $10$) complete bipartite pattern. There will be $\binom{20}{10}$ many instances of the smaller pattern inside the larger pattern, even if there are only 30 different nodes across these matches. A second challenge is scalability. The application would like to search the patterns ideally across 100s of billions of edges, by generating the graph from a much larger time period of transactions.

\customheader{Other Patterns:} We omit a detailed description but the interviewee described two other use cases:
\begin{squishedlist}
\item The first application detects gambling activity on Alipay, which is an online payment platform. The input graph contains customers' Alipay accounts as nodes and {\em Alipay groups}, a service to allow groups of people to exchange money amongst themselves. The edges are \texttt{membership} edges between customers and groups. For each gambling game, a set of gamblers start and join a new group. Similar to the fake transactions application, the pattern in its simplified form forms a near-complete bipartite graph of accounts and groups.
\item Another application monitors attacks and threats on Alibaba Cloud's  network and traffic graph, which contains information about the hosts, e.g., IP addresses, ports, domain names, and the traffic between the hosts. The application searches a star pattern consisting of a single node with several labeled edges, some with regex patterns, to match address and host name patterns. The graph is highly dynamic, and the application searches patterns on a snapshot that contains only a few days' data, which contains over 100 billion edges.
\end{squishedlist}

\subsubsection{Application at a Large Financial Institution}
Our consultant interviewee described a fraud detection application for a large financial institution that had customer accounts and different transactions between accounts.   The searched pattern was quite complex and was not described in detail. Broadly it involves searching for connected accounts over very long paths in the graph, where the nodes that are close to the center of these paths have an unusually high number of transactions, i.e., edges, and amounts of transactions. Interestingly, when asked how they observed that this pattern is a signal of fraud, the interviewee said that initially he manually searched for fraudulent patterns. Specifically, he visualized large chunks of the graph, sometimes containing several  thousands of nodes, on a visualization software, and eye-balled known fraudsters' activities. He noticed this pattern as part of this visualization activity.

%% file: 05c_alibaba_amazon_question_answering.tex
\subsection{Question Answering with Personal Assistant Products of Alibaba and Amazon}
Alibaba and Amazon both produce voice-controlled personal assistant products, {\em AliGenie}~\cite{aligenie} and {\em Alexa}~\cite{alexa}, respectively, that can be accessed from different devices, such as smart speakers or mobile phones for several services. One of these services is to answer factual questions asked by human users through speech.
Our interviewees from both companies described similar applications that use a knowledge graph to answer these questions. 

The questions asked by users can be highly varied and require knowledge from public information, corporate information, or user-specific information, e.g., about the movies the user has seen. 
Our interviewees both described similar applications that use a knowledge graph to answer questions.

In both cases, the interviewees could not provide the details of these graphs but briefly mentioned that the graph used to answer these questions include many internal and external sources. In Alibaba's case, the Chinese edition of Wikipedia, information from Alibaba Music, information about the businesses that sell products on Alibaba were mentioned. Both interviewees mentioned the  challenges of integrating such numerous and diverse internal and external sources.

The high-level steps of both applications were very similar and consisted of components that perform voice recognition and natural language processing to understand the important entities used in the query. For example, in the question ``What are the movies that Tom Hanks played in 2018?'', ``Tom Hanks'' would be identified as the main entity. Then all nodes that are referred to as Tom Hanks are identified from the graph and a local search is made around these nodes to find nodes that are of type movies and have date information. The details of these searches were not described but in both cases many nodes will be matched, returned, and ranked before an answer is produced. Both interviewees mentioned doing semantic inference using ontologies, e.g., to infer that the word ``played" is semantically related to ``acted in'', to extend the search or rank the results.

Interviewees provided few details about the software on which the knowledge graph is stored and the search is performed. In the case of Amazon, the graph was stored in RDF format and indexed in an in-house software (not an RDF system).  In the case of Alibaba, although each edge (or fact) in the graph was extracted as an RDF triple, the graph was then stored in an in-house graph database that supports the property graph model.

%% file: 05d_contingency_analysis.tex
\subsection{Contingency Analysis of Power Failures at State Grid}%
\label{sec:contingency-analysis}

Contingency analysis is a preemptive analysis
done on an electric power grid to check the severity of different possible failures.

Our interviewee from StateGrid described a contingency analysis system designed for
the grid in one Chinese province. In contrast to other applications which often used one large graph, this application, logically, uses a very large number of small graphs. Interestingly, these graphs are very similar to each other and the application repeats the same computation on each graph in parallel. We describe the input graph, the computation, and the software used by the application.

\customheader{Input Graph:} The application has a {\em base graph} that represents the components of the power grid using the abstract \emph{bus-branch} standardized model~\cite{bus-branch-model}:

\begin{squishedlist}
    \item Vertices correspond to buses that represent electrical nodes, which
    can include power system elements like substations, loads, and generators.
    Operational parameters such as bus id, load power, voltage magnitude,
    voltage angle, self impedance, and power injection are stored as vertex
    attributes. There are approximately 2.5K vertices.

    \item Edges correspond to branches that represent electrical paths for
    current flows, such as transmission lines and transformers. Operational
    parameters such as power flow, line impedance, and transformer turns ratio
    are stored as edge attributes. There are approximately 3K edges.
\end{squishedlist}

\noindent This is a dynamic graph and the attributes on the edges are changing every few seconds and the topology changes, e.g., a new node is added or removed, every few minutes.

\customheader{Computation:}
To determine how the failure of a component affects the flow of power in the
grid, the application generates a few thousand logical {\em derived graphs} from the base graph. Each derived graph modifies the base graph slightly, say by removing a single edge, to simulate a potential failure. For each derived graph $G'$, the application formulates some power equations. We do not provide the details of these equations but an overview can be found in reference~\cite{contingency-analysis}.  In a simplified form, readers can think of these as equations of the form $Ax = B$, where $A$ and $B$ are power-related matrices, and each row of which represents information about the neighborhood of a vertex in the derived graph. These equations are solved in parallel using matrix factorization. We note that there is a significant potential to reuse the computation results across the derived graphs, as the graphs are very similar. The solution $x$'s are analyzed to assign severity values to each derived graph and an alert is raised for abnormally high severity values, indicating the system has found a potential failure case, which could have severe outcomes.

\customheader{Software used:}
The base graph is stored in Tigergraph\cite{tigergraph}. Derived graphs are logical and not explicitly stored but their corresponding matrices A and B are read from Tigergraph in parallel and moved to a custom code that solves the power flow linear algebra equations. The overall latency of the application is several seconds. Although this is not done currently, some equations can also be directly solved using iterative vertex-centric computations on the base graph directly~\cite{contingency-analysis}.

%% file: 06endtext.tex
\section{Related Work}\label{sec:related-work}

\noindent To the best of our knowledge, our survey is the first study that has
been conducted across users and of a wide spectrum of graph technologies and various public information about these technologies to understand graph datasets, computations, and software that is in use, the business applications that use graphs and the challenges users face.

Several surveys in the literature have conducted user studies to compare the
effectiveness of different techniques used to perform a particular graph
processing task, primarily in visualization~\cite{Holten09auser, Bridgeman2001}
and query languages~\cite{DBLP:conf/icde/JayaramKLYE16, Rath2017AreGQ,
    Pienta:2016:VIV:2909132.2909246}. Additionally, several software vendors
have conducted surveys of their users to understand how their software is used
to process graphs. Some of these surveys are publicly
available~\cite{neo4j-survey,flink-survey,spark-survey}. However, these surveys
are limited to studying a specific software product.

There are also numerous surveys in the literature studying different topics
related to graph processing. Examples include surveys on query languages for
graph database systems and RDF engines~\cite{AnglesABHRV17,
    holzschuher2013performance, Haase2004}, graph
algorithms~\cite{Aggarwal2010,CGF:CGF12800, herman2000graph,
    katifori2007ontology}, graph processing systems~\cite{Batarfi2015,
    Lu:2014:LDG:2735508.2735517}, and visualization~\cite{CGF:CGF12872,
    cui2007survey}. These surveys do not study how users use the technologies in
practice.

\section{Conclusion and Future Work}\label{sec:conclusions}

\noindent Managing and processing graph data is prevalent across a wide range of
fields in research and industry. We surveyed 89 users, interviewed 8 users, and reviewed user emails, code repositories, and whitepapers of a large suite of software products. The participants' responses and
our review provide useful insights into the types of graphs users have, the
software and computations users use, the business applications they develop, and the major challenges users face when
processing their graphs. We hope that these insights and in particular the challenges we highlight will help guide research on graph processing.

We conclude with two final remarks. First, we found product-order-transaction
graphs to be the most popular type of graph. Workloads that process these
product data appear in popular SQL benchmarks, such as
TPC-C~\cite{tpc-benchmarks}, and are well studied in research on relational
systems. However, several existing graph benchmarks, such as LDBC~\cite{ldbc-benchmarks}
and Graph500~\cite{graph500-benchmarks}, do not yet provide workloads and data
to process product graphs. One such benchmark is the WatDiv benchmark~\cite{aluc:watdiv} that generates RDF triples containing information about products and purchases.  Developing similar benchmarks and popularizing their use would be highly beneficial to the research community. Such benchmarks are great facilitators of research.

Second, query languages and APIs emerged as one of the top challenges in our
survey and certainly the most popular discussion topic in emails and code
repositories. These challenges can be partly mitigated by a collaborative effort
to standardize the query languages of different graph software that satisfy
users' needs. One such successful effort is the adoption of SPARQL as a standard
for querying RDF data. Similar efforts are ongoing for developing standard query
languages and JDBC-like interfaces~\cite{jdbc} for property graphs, such as the
Gremlin language~\cite{DBLP:journals/corr/Rodriguez15} and the efforts to combine openCypher~\cite{opencypher}, PGQL~\cite{pgql}, and G-CORE~\cite{AnglesABBFGLPPS18} to create GSQL~\cite{gsql}. There is also ongoing
effort to develop a standard set of linear algebra operations for expressing
graph algorithms~\cite{6670338}.

\section{Acknowledgments}
\noindent We are grateful to Chen Zou for helping us in using online survey
tools and drafting an early version of this survey. We are also grateful to
Nafisa Anzum, Jeremy Chen, Pranjal Gupta, Chathura Kankanamge, and Shahid Khaliq for their valuable comments on the survey
and help in categorizing the academic publications, user emails, and issues. We would like to thank our online survey participants and our interviewees: Brad Bebee, Scott Brave, Jordan Crombie, Luna Dong, William Hayes, Thomas Hubauer, Peter Lawrence, Stephen Ruppert,  Chen Yuan,  with a special thanks to Zhengping Qian for the many hours of follow-up discussions after our interview. Finally, we would like to thank the anonymous reviewers for their valuable comments. This research was partially supported by multiple Discovery Grants from Natural Sciences and Engineering Council (NSERC) of Canada.

%% file: 07appendix.tex
\begin{appendix}

    \section{Choices of Graph
        Computations}\label{app:methodology-graph-computations}

    \noindent One way to ask this question is to include a short-answer question
    that asks ``What queries and graph computations do you perform on your
    graphs?'' However, the terms graph queries and computations are very general
    and we thought this version of the question could be under-specified. We
    also knew that participants respond less to short-answer questions, so
    instead we first asked a multiple choice question followed by a short answer
    question for computations that may not have appeared in the first question
    as a choice.

    In a multiple choice question, it is very challenging to provide a list of
    graph queries and computations from which participants can select, as there
    is no consensus on what constitutes a graph computation, let alone a
    reasonable taxonomy of graph computations. We decided to select a list of
    graph queries and computations that appeared in the publications of six
    conferences, as described in Section~\ref{sec:academic-publications-review}.
    We use the term graph computation here to refer to a query, a problem, or an
    algorithm.

    For each of the 90 papers, we identified each graph computation, if (i)~it
    was directly studied in the paper; or (ii)~for papers describing a software,
    it was used to evaluate the software. We used our best judgment to
    categorize the computations that were variants of each other or appeared as
    different names under a single category. For example, we identified motif
    finding, subgraph finding, and subgraph matching as \emph{subgraph
        matching}. When reviewing papers studying linear algebra operations,
    e.g., a matrix-vector multiplication, for solving a graph problem such as
    BFS traversal, we identified the graph problem and not the linear algebra
    operation as a computation.

    Finally, for each identified and categorized computation, we counted the
    number of papers that study it and selected the ones that appeared in at
    least 2 papers. In the end, we provided the participants with the 13 choices
    that are shown in Table~\ref{table:graph-computations}.

    \vspace{-0.5em}
    \section{Choices of Machine Learning
        Computations}\label{app:methodology-ml-computations}
    \vspace{-0.1em}

    \noindent Similar to graph computation, machine learning computation is a
    very general term. Instead of providing a list of ad-hoc computations as
    choices, we reviewed each machine learning computation that appeared in the
    90 graph papers we had selected. Specifically, the list of machine learning
    computations we identified included the following: (i)~\emph{high-level
        classes of machine learning techniques}, such as clustering,
    classification, and regression; (ii)~\emph{specific algorithms and
        techniques}, such as stochastic gradient descent and alternating least
    squares that can be used as part of multiple higher-level techniques; and
    (iii)~\emph{problems} that are commonly solved using a machine learning
    technique, such as community detection, link prediction, and
    recommendations. We then selected the computations, i.e., high-level
    techniques, specific techniques, or problems, that appeared in at least 2
    papers. As in the graph computations question, we used our best judgment to
    identify and categorize similar computations under the same name.

    \vspace{-0.5em}
    \section{Storage in Multiple Formats}\label{app:dataformats}
    \vspace{-0.1em}
    \noindent We asked the 33 participants who said that they store their data
    in multiple formats, which formats they use as a short-answer question. Out
    of the 33 participants, 25 responded. Their responses contained explicit
    data storage formats as well as the internal formats of different software.
    Table~\ref{table:count-storage-formats} shows the number of responses we
    received for the main formats. A relational database and a graph database
    format combination was the most popular combination. Other combinations
    varied significantly, examples of which include HBase and Hive, GraphML and
    CSV, and XML and triplestore.

\begin{table}[t!]
    \centering
    \caption{Data storage formats.}%
    \label{table:count-storage-formats}
    \iftablesqueeze\vspace{-0.7em}\fi
    \begin{tabular}{p{5cm}r}
      \toprule
      \multicolumn{1}{l}{\textbf{Data Storage Format}} & \textbf{\#} \\ \midrule
      Graph Databases                                    & 10          \\
      Relational Databases                               & 8           \\
      RDF Store                                          & 5           \\
      NoSQL Store (Key-value, HBase)                     & 5           \\
      XML / JSON                                         & 4           \\
      JGF / GML / GraphML                                & 4           \\
      CSV / Text files                                   & 3           \\
      Elasticsearch                                      & 3           \\
      Binary                                             & 2           \\ \bottomrule
    \end{tabular}
\end{table}

\begin{table}[t!]
    \centering
    \caption{Graph sizes in user emails and issues.}%
    \label{table:sizes-email}
    \iftablesqueeze\vspace{-0.7em}\fi
    \begin{subtable}[b]{.4\columnwidth}
        \centering
        \subcaption{Number of vertices.}%
        \label{table:no-of-vertices-2}
        \iftablesqueeze\vspace{-0.3em}\fi
        \begin{tabular}{lr}
          \toprule
          \textbf{Vertices} & \textbf{\#} \\ \midrule
          100M$-$1B         & 10          \\
          1B-10B            & 17          \\
          10B$-$100B        & 1           \\
          \textgreater100B  & 2           \\ \bottomrule
        \end{tabular}
    \end{subtable}
    \begin{subtable}[b]{.4\columnwidth}
        \centering
        \subcaption{Number of edges.}%
        \label{table:no-of-edges-2}
        \iftablesqueeze\vspace{-0.3em}\fi
        \begin{tabular}{lr}
          \toprule
          \textbf{Edges}   & \textbf{\#} \\ \midrule
          1B$-$10B         & 42          \\
          10B$-$100B       & 17          \\
          100B$-$500B      & 6           \\
          \textgreater500B & 1           \\ \bottomrule
        \end{tabular}
    \end{subtable}
\end{table}

\begin{table}[t!]
    \centering
    \caption{Challenges found in user emails and issues.}%
    \label{table:email-challenges}
    \iftablesqueeze\vspace{-0.7em}\fi
    \begin{tabular}{lr}
      \toprule
      \multicolumn{1}{l}{\textbf{Challenge}} & \textbf{\#}   \\ \midrule
      \multicolumn{2}{c}{\textbf{Graph DBs and RDF Engines}} \\ \midrule
      High-degree Vertices                   & 24            \\
      Hyperedges                             & 18            \\
      Triggers                               & 18            \\
      Versioning and Historical Analysis     & 14            \\
      Schema \& Constraints                  & 10            \\ \midrule
      \multicolumn{2}{c}{\textbf{Visualization Software}}    \\ \midrule
      Layout                                 & 31            \\
      Customizability                        & 30            \\
      Large-graph Visualization              & 8             \\
      Dynamic Graph Visualization            & 4             \\ \midrule
      \multicolumn{2}{c}{\textbf{Query Languages}}           \\ \midrule
      Subqueries                             & 7             \\
      Querying Across Multiple Graphs        & 6             \\ \midrule
      \multicolumn{2}{c}{\textbf{DGPS and Graph Libraries}}  \\ \midrule
      Off-the-shelf Algorithms               & 41            \\
      Graph Generators                       & 7             \\
      GPU Support                            & 3             \\ \bottomrule
    \end{tabular}
\vspace{-1em}
\end{table}

    \vspace{-0.2em}
\section{Other Tables from the Survey}\label{app:other-tables}
\noindent Table~\ref{table:sizes-email} shows the sizes of graphs we found in
user emails and issues. Table~\ref{table:email-challenges} shows the number of
emails and issues we identified for each specific challenge we discussed in
Section~\ref{sec:email-challenges}. Table~\ref{table:counts-emails-issues} shows
the total number of emails and issues we reviewed for each software product from
January to September of 2017. The table also shows the number of commits in the
code repositories of each software product during the same period.

\begin{squishedtable}
    \centering
    \caption{The number of reviewed emails and issues, and the code commits
        in the repositories of each software product.}%
    \label{table:counts-emails-issues}
    \setlength\tabcolsep{0.25em}
    \iftablesqueeze\vspace{-0.7em}\fi
    \begin{tabular}{p{2.3cm}p{1.9cm}rrr}
      \toprule
      \multicolumn{1}{l}{\textbf{Technology}}                              & \multicolumn{1}{l}{\textbf{Software}} & \textbf{\#Emails} & \textbf{\#Issues} & \textbf{\#Commits} \\ \midrule
      \multirow{9}{*}{\parbox{2cm}{Graph Database System}}                 & ArrangoDB                             & 140               & 466               & 5264               \\ \cmidrule{2-5}
                                                                           & Caley                                 & 50                & 57                & 151                \\ \cmidrule{2-5}
                                                                           & DGraph                                & 175               & 558               & 760                \\ \cmidrule{2-5}
                                                                           & JanusGraph                            & 225               & 308               & 411                \\ \cmidrule{2-5}
                                                                           & Neo4j                                 & 286               & 243               & 4467               \\ \cmidrule{2-5}
                                                                           & OrientDB                              & 169               & 668               & 918                \\ \cmidrule{2-5}
                                                                           & Sparksee                              & 8                 & NA                & NA                 \\ \midrule
      \multirow{3}{*}{RDF Engine}                                          & Apache Jena                           & 307               & 126               & 471                \\ \cmidrule{2-5}
                                                                           & Virtuoso                              & 72                & 61                & 179                \\ \midrule
      \multirow{6}{*}{\parbox{2.2cm}{Distributed Graph Processing Engine}} & Apache Flink (Gelly)                  & 34                & 68                & 48                 \\ \cmidrule{2-5}
                                                                           & Apache Giraph                         & 19                & 34                & 23                 \\ \cmidrule{2-5}
                                                                           & Apache Spark (GraphX)                 & 23                & 28                & 11                 \\ \midrule
      Query Language                                                       & Gremlin                               & 409               & 206               & 1285               \\ \midrule
      \multirow{7}{*}{Graph Library}                                       & Graph for Scala                       & 10                & 12                & 18                 \\ \cmidrule{2-5}
                                                                           & GraphStream                           & 18                & 26                & 7                  \\ \cmidrule{2-5}
                                                                           & Graphtool                             & 121               & 66                & 172                \\ \cmidrule{2-5}
                                                                           & NetworKit                             & 37                & 30                & 236                \\ \cmidrule{2-5}
                                                                           & NetworkX                              & 78                & 148               & 171                \\ \cmidrule{2-5}
                                                                           & SNAP                                  & 57                & 17                & 34                 \\ \midrule
      \multirow{6}{*}{Graph Visualization}                                 & Cytoscape                             & 388               & 264               & 8                  \\ \cmidrule{2-5}
                                                                           & Elasticsearch X-Pack Graph            & 50                & 38                & NA                 \\ \cmidrule{2-5}
                                                                           & Gephi                                 & NA                & 147               & 10                 \\ \cmidrule{2-5}
                                                                           & Graphviz                              & NA                & 58                & 277                \\ \midrule
      Graph Representation                                                 & Conceptual Graphs                     & 30                & NA                & NA                 \\ \bottomrule
    \end{tabular}
\vspace{-1.3em}
\end{squishedtable}

\input{05e_other_applications}

\end{appendix}

%% file: 05e_other_applications.tex
\vspace{-0.2em}
\section{Other Applications from Interviews}%
\label{appendix:other-interviews}

\vspace{-0.4em}
\customheader{Large Scale Data Integration for Analysis of Turbines:}
Our interviewee from Siemens described an application that Siemens engineers use to analyze different properties of gas turbines Siemens produces using a knowledge graph. The application emphasized the advantage of using graphs to integrate different sources of corporate data, in this case mainly about where turbines are installed, measurements from the installed turbines' sensors, and information about maintenance activity on the turbines. The knowledge graph is stored in an RDF engine and engineers ask queries, such as ``What is the mean time failure of turbines with coating loss?'' through a visual interface where they navigate the different types of nodes in the knowledge graph and express aggregations. The visually expressed queries get translated to SPARQL queries.

\vspace{-0.3em}
\customheader{Contact Deduplication:} One of our interviews was with two engineers from a  a company called FullContact that manages contact information about individuals by integrating public and manually curated information, which is sold to other businesses. An over 10B-edge and 4B-vertex graph models this contact information as follows: nodes represent different pieces of information, such as addresses, phone numbers, and edges between nodes indicate the likelihood that the information belongs to the same individual. The company uses GraphX to run a connected components-like algorithm to finding the contacts that are likely to belong to the same individual.

\vspace{-0.3em}
\customheader{Other applications using knowledge graphs:}  One of our interviewees was a consultant to a chemical company specializing in agricultural chemicals. The company has an over 30 billion-edge knowledge graph on pesticides, seeds, chemicals that is stored in a commercial RDF system. This graph is used by many applications, such as,  to track the evolution of seeds,  to power internal wiki pages and tools used by analysts.
Another interviewee was the founder of a startup that works on tools that can be used by biologists to publish biological knowledge. Our interviewee described examples of knowledge graphs that model the cellular activity in the context of different species' different tissues. Triples included facts about which genes transcribe which protein and which proteins' presence decreases other proteins' presence, etc.~\cite{openbel}. The example applications were similar broadly to our interviewee from the chemical company and  power wikis and website which biologists use to analyze these interactions.